\documentclass[aps,prb,twocolumn,floats]{revtex4-2}
\usepackage{tikz}
\usepackage{epsfig}
\usepackage{amsmath, amssymb}
\usepackage{graphicx}
\usepackage{soul}
\usepackage{braket}
\usepackage[colorlinks=true,linkcolor=blue]{hyperref}
\usepackage{subfigure}
\usepackage{color}

\newcommand{\blue}{\textcolor{blue}}

\begin{document}
\title{Transitions and Critical Divergences in Periodically Hopping Modulated Su-Schrieffer-Heeger Chains}
\author{Surajit Mandal$^{1,2}$}
\email{surajitmandalju@gmail.com}
\author{Satyaki Kar$^{2}$}
\email{satyaki.phys@gmail.com}
\affiliation{$^{1}$Department of Physics, Jadavpur University, Kolkata, West Bengal -700032, India\\$^{2}$Department of Physics, AKPC Mahavidyalaya, Bengai, West Bengal -712611, India}

\begin{abstract}
We use a curvature renormalization group (CRG) approach to study the topological phase transitions in a Su-Schrieffer-Heeger chain and its extensions coming from periodic hoppinng modulations. A curvature function is defined in terms of system parameters near high-symmetry points where the divergence of this function at critical points, in analogy to usual phase transitions, signals a topological phase transition. According to this theory, the phase transition line for the two-site  Su-Schrieffer-Heeger (SSH) model is visible at the critical line $\Delta=0$ where the curvature function diverges. Our study involves this model and also the modulated one with periodicity of four lattice spacing where the curvature function not only diverges at the topological phase transition point (Dirac-like) $|\Delta/t|=\sqrt{2}$ but also shows faster divergence at the non-topological gapless point $\Delta=0$. We further notice faster divergence of correlation length for $\Delta\rightarrow0$ as compared to that for the $|\Delta/t|\rightarrow\sqrt{2}$ resulting in two different sets of critical exponents making them lie in different universality classes. The edge state exhibits very slow decay into the bulk near the $\Delta=0$ point while a much quicker decay from edge into bulk is discernible around the $|\Delta/t|=\sqrt{2}$ point. We also continue similar analysis for a SSH model with hopping periodicity of eight lattice spacings.
\end{abstract}
\maketitle
\section{Introduction}\label{sec1}

A topological order is non-local in nature which is usually quantified by an integer-valued order parameter called topological invariant. These are often calculated from the momentum space integration of a function called curvature function, whose
precise form depends on the dimension and the symmetry class of the system considered\cite{ludwig,kitaev}.
Although topology is a global property of the entire manifold of the Brillouin zone (BZ), the knowledge about topology can be entirely encoded in the curvature function near a high-symmetry point (HSP).
In topological phase transitions (TPTs),
the topological invariant jumps discretely as the curvature function usually diverges at certain HSPs in momentum space with the sign changing across the point of TPT. The curvature renormalization group (CRG) method can capture the TPTs solely based on the renormalization of the curvature function near the HSP\cite{chen0} in any noninteracting\cite{rg,rao}, interacting\cite{rg5} or say, periodically driven\cite{rg2} systems.

In this paper, we consider a periodically hopping modulated version of the Su-Schrieffer-Heeger chain - a celebrated non-interacting model for studying the TPTs in one-dimensional (1D) condensed matter systems and discuss the nature of transitions and topological states in the light of CRG analysis.

The paper is organized as follows. In Sec.\ref{sec2} we briefly discuss the curvature renormalization methodology while Sec.\ref{sec3} details the formulation particularly for the SSH(like) model with hopping modulation of different periodicities. It also provides a detailed analysis of symmetry, topological invariant calculations as well as the features of spectra and end states. Finally, we summarize our results in Sec.\ref{sec4} and discuss its novelty and possible future directions of work.

\section{Curvature Renormalization}\label{sec2}
In this section, we briefly discuss the curvature renormalization group method previously outlined in Refs.\cite{rg,rg1,rg2,rg3,rg4,rg5,rg6,edge1}. As we already mentioned, one can calculate the topological invariant in a topological system by integrating
a function over the Brillouin zone called the curvature function (depending the
dimensions and the underlying symmetry class of the system\cite{rg}, it can either be the Berry connection, the Berry curvature, or the
Pfaffian of an appropriate “sewing matrix”). More precisely, the topological invariant $C$ is given as
\begin{equation}\label{curvature1}
 C=\int_{BZ}\frac{d^{D}k}{(2\pi)^{D}}F(k,{\bf M}),
\end{equation}
in which $F(k,{\bf M})$ and ${\bf M} =(M_{1},M_{2},...,M_{i},...)$ are, respectively, the curvature function and the set of all the tuning parameters in the Hamiltonian. 
Different $C$ values represent different phases separated by topological phase transitions (TPTs).
As studied in Refs.\cite{rg,rg4,edge1}, $F(k,{\bf M})$ displays a peak at a high-symmetry point (HSP) $k_{0}$ obeying the relation $F(k_{0}+\delta k,{\bf M})=F(k_{0},{\bf M}^{\prime})$ for small $\delta k$\cite{rg4} where for an inversion symmetric system, one can use the Ornstein-Zernike Lorentzian form\cite{rg1}
\begin{equation}\label{curvature2}
 F(k_{0}+\delta k,{\bf M})=\frac{F(k_{0},{\bf M})}{1+\xi_{k_{0}}^2\delta k^2},
\end{equation}
where $\xi_{k_{0}}$ is the correlation length. The system undergoes a phase transition at the critical point ${\bf M_{c}}$ (say) where the gap closes at HSP $k_{0}$ in the Brillouin zone. The length scale $\xi_{k_{0}}\rightarrow\infty$ for $M\rightarrow M_{c}$ resulting in narrowing of the Lorentzian of Eq. (\ref{curvature2}). This makes the curvature function to diverge when approaching the critical point  $M\rightarrow M_{c}$ and more specifically,
\begin{equation}\label{curvature3}
\lim_{M\rightarrow M_{c}^{+}} F(k_{0},{\bf M})=-\lim_{M\rightarrow M_{c}^{-}} F(k_{0},{\bf M})=\pm\infty,
\end{equation}
where ${\bf M_{c}^{+}}$ and $ {\bf M_{c}^{-}}$ indicate the two sides of the phase boundary at $\bf{M = M_c}$ in the parameter regime. 
This blowing up of $F(k,{\bf M})$ as one approaches a HSP: $k\rightarrow k_{0}$ with a flipped sign on either side of the TPT at  ${\bf M=M_c}$ is compatible with a discrete jump in the value of topological invarint at TPT.

We expect divergent behavior close to the TPT as:
\begin{equation}\label{curvature3}
 F(k_{0},{\bf M})\propto |{\bf M}-{\bf M_{c}}|^{-\gamma},~~\xi_{k_{0}}\propto |{\bf M}-{\bf M_{c}}|^{-\nu},
\end{equation}
with exponents $\gamma$ and $\nu$ denoting the critical exponents characterizing the TPT. These critical exponents are not independent but follow a scaling law $\gamma=\nu$ due to conservation of topological invariant for ${\bf M}\rightarrow {\bf M_{c}}$\cite{edge1} without crossing ${\bf M_c}$.

As mentioned before, the system undergoes a topological phase transition at ${\bf M}={\bf M_{c}}$ where the topological invariant $C$ changes and the bulk band gap vanishes at HSP $k_{0}$ where the curvature function diverges. Thus the CRG method can be summerized by demanding that at $k_{0}+\delta k$ for a parameter set ${\bf M}$, one need to search for a new set ${\bf M^{\prime}}$ that satisfy
\begin{equation}\label{curvature4}
F(k_{0}+\delta k,{\bf M})=F(k_{0},{\bf M^{\prime}}).
\end{equation}
 Due to this process, the divergence of $F(k,{\bf M})$ reduces at $k_{0}$ which is familiar as deviation-reduction mechanism\cite{rg}. Defining $dM_{i}=M_{i}^{\prime}-M_{i}$ and $\delta k^2=dl$ and expanding Eq. (\ref{curvature4}) to leading order gives the RG equation for parameter ${\bf M}$ as:
\begin{equation}\label{curvature5}
\frac{dM_{i}}{dl}=\frac{1}{2}\frac{\partial_{k}^2F(k,{\bf M})\Big|_{k=k_{0}}}{\partial_{M_{i}}F(k_{0},{\bf M})}.
\end{equation}
In this CRG procedure, the critical point can be obtained when $|\frac{d{\bf M}}{dl}|\rightarrow\infty$ (in which flow directs away) and stable (unstable) fixed point can be estimated by $|\frac{d{\bf M}}{dl}|\rightarrow 0$ in which flow directs into (away)\cite{rg7}. In order to investigate the TPT, this CRG procedure will now be applied to our periodically hopping modulated SSH chain.

\section{SSH chain with periodic hopping modulations}\label{sec3}

The Su-Schrieffer-Heeger (SSH) model \cite{ssh,wall2}, introduced in the context of polyacetylene is given by a one-dimensional tight-binding Hamiltonian. The SSH model for $L=M*N$ (where $M$ and $N$ denote the number of sublattices and unit cells, respectively) sites with staggered nearest-neighbor hopping defined as\cite{kar,mandal}
\begin{equation}\label{1}
  H_{SSH}=\sum_{i}^{L-1}(t+\delta_{i})c_{i}^{\dagger}c_{i+1}+h.c
\end{equation}
here $c_{i}^{\dagger}$( $c_{i}$) is the electron creation (annihilation) operator and the periodic modulation in nearest-neighbor hopping strength $t$ is obtained by $\delta_{i}=\Delta \cos[(i-1)\theta]$ with $i=1,2,3,......, N$. In general, we get $\delta_{i+1}=\Delta \cos(\frac{2\pi i}{M})$ with $\theta=2\pi/M$ and the chain is represented by a $M\times M$ Hamiltonian matrix having $M$ number of eigenmodes. The Hamiltonian shows chiral or sublattice symmetry, and again, the transformation of $c_{i}\rightarrow(-1)^ic_{i}^\dagger$ makes
$H_{SSH}\rightarrow H_{SSH}$, also indicating the sublattice or chiral symmetry\cite{mandal}. Importantly, the chiral symmetry requires the total number of sites to be even when we consider a model under periodic boundary conditions (PBC)\cite{mandal}.

\subsection{Case I: $\theta=\pi$}
This case corroborates the original SSH model\cite{ssh,wall2} and was studied recently in the context of edge state behavior in Ref.\cite{kar}. Here, the size of the first Brillouin zone (FBZ) lies within the interval $k\in [-\pi/2,\pi/2]$ with reciprocal vector becoming $G=\pi$. Therefore, one can find a gap closing high-symmetry point (HSP) at $k_{0}=0$ and $k_{0}=\pm \pi/2$. Among them $k_{0}=0$ HSP is not what we are mainly interested with as we discuss below.
\subsubsection{Hamiltonian and the phase diagram} 
\begin{figure}
   \vskip -.4 in
   \begin{picture}(100,100)
     \put(-70,0){
  \includegraphics[width=.45\linewidth, height=1.25 in]{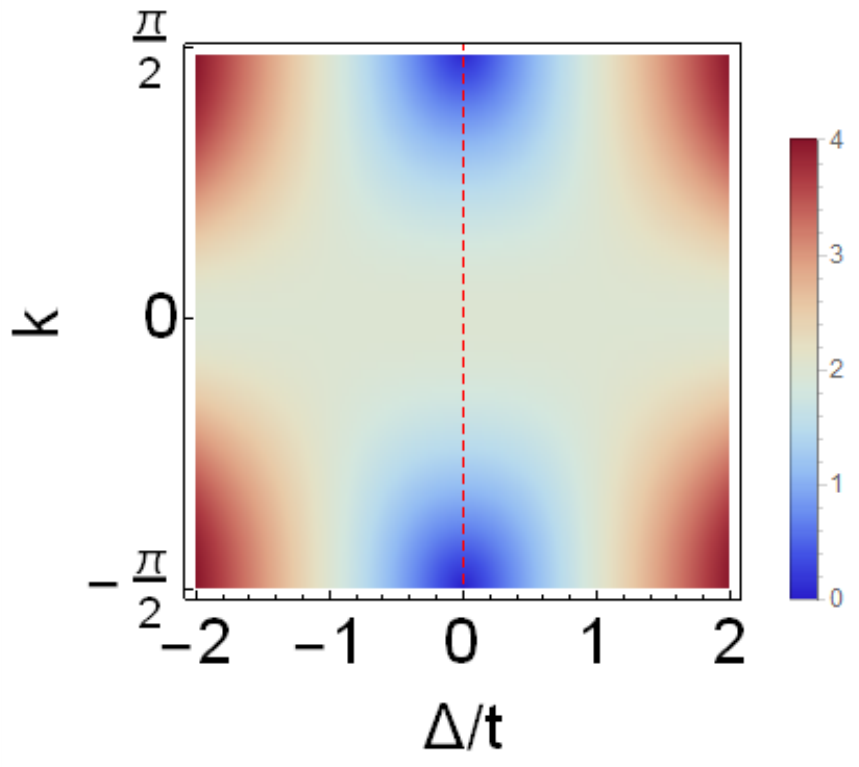}
  \includegraphics[width=.45\linewidth, height=1.25 in]{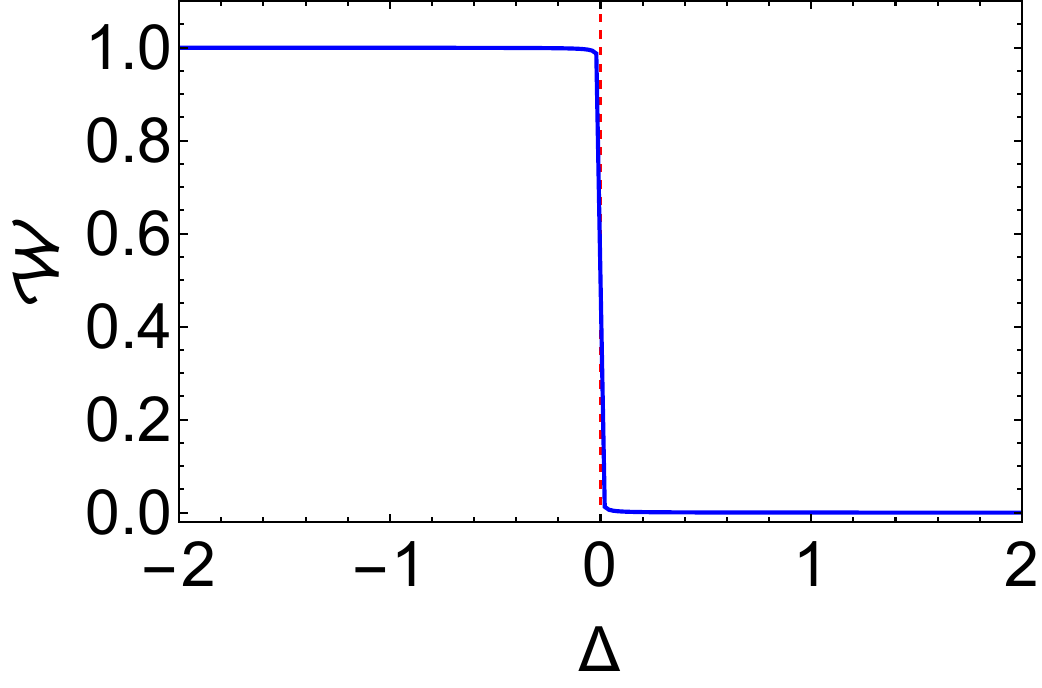}}
     \put(-10,62){(a)}
     \put(125,58){(b)}
    \end{picture}\\
     \vskip -.00004 in
   \begin{picture}(100,100)
     \put(-10,0){
       \includegraphics[width=.48\linewidth,height=1.25 in]{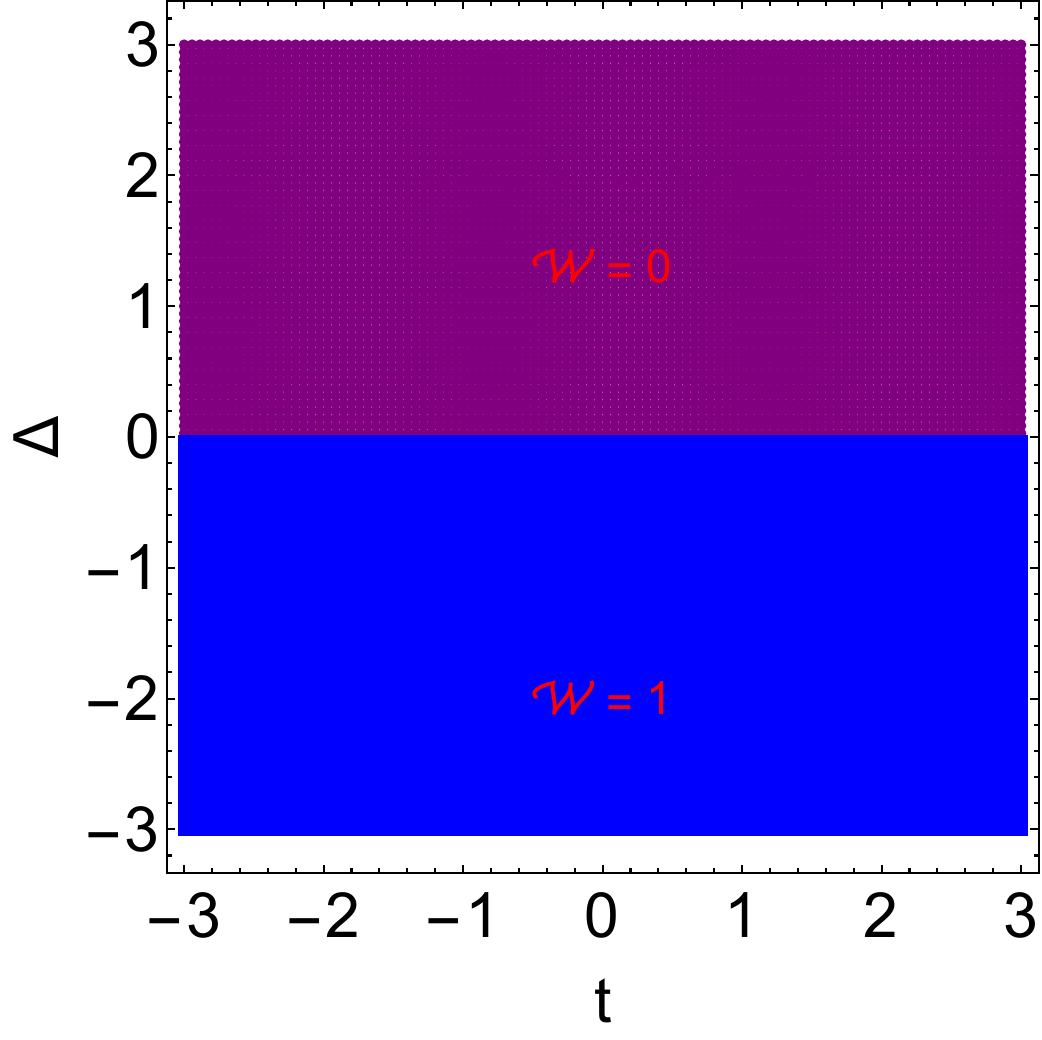}}
   \put(-10,58){(c)}
   \end{picture}
  \vskip -0.1 in
\caption{(a) The bulk gap in the $(\Delta/t-k)$-plane for simple two-site SSH chain. $\Delta=0$ is the gap closing line when $k=\pm\pi/2$. The deep blue in the color bar indicates gap closing (b) The winding number as a function of $\Delta/t$. (c) The phase diagram showing the topological (blue region) and trivial (purple) region.} 
\label{fig2}
\end{figure}

The single-particle Bloch Hamiltonian in momentum space takes the following form
\begin{equation}\label{5}
H_{k}=
\begin{pmatrix}
0 & P_{k} \\
 P_{-k} & 0
\end{pmatrix},
\end{equation}
where, $P_{k}=(t+\Delta)+ (t-\Delta)e^{-2ik}$. It can be rewritten using the Pauli matrices $\{\sigma_{i}\}$ notation as
\begin{equation}\label{5a}
H_{k}=\overrightarrow{h(k)}.\sigma
\end{equation}
where
\begin{eqnarray}\label{5b}
h_{x}(k)&=&(t+\Delta)+(t-\Delta)\cos(2k)\nonumber\\
h_{y}(k)&=&(t-\Delta)\sin(2k)\nonumber\\
h_{z}(k)&=&0
\end{eqnarray}
In the topological tenfold way classification scheme, the model belongs to the BDI class of universality\cite{s1,altand}. The energy eigenvalues are given by
\begin{equation}\label{6}
E(k)_{\pm}=\pm\sqrt{2}\sqrt{t^2\cos^2k+\Delta^2\sin^2k}
\end{equation} 
which exhibits two bands in the energy spectrum. As soon as we consider the system under PBC, the band gap closes at the edge of FBZ at $k=k_0=\pm\pi/2$ for $\Delta=0$. Though the gap also closes at $k=k_0=0$ for $t=0$, we avoid that as all energy scales are defined in terms of $t$ and we generally don't consider $t=0$\cite {rg} (notice Fig. \ref{fig2}(a)). The finite band gap at the boundary of FBZ is noticeable for $k=\pm\pi/2$ for $\Delta\ne 0$ and has a magnitude of $E_{gap}\sim|\Delta|$\cite{comment}. Thus, here only $k_0=\pm\pi/2$ represents the gap closing HSP making it a topological phase transition point (TPT) for $\Delta=0$\cite{mandal,kar}. The dispersion remains linear ($i.e.,$ Dirac-like) near this point and two zero-energy modes (ZES) are obtained in the topological regime\cite{mandal,bernevig}. 

The topological phases are, as usual, characterized by bulk topological invariant, namely the winding number $\mathcal{W}$ given as\cite{w1,w2}
\begin{equation}\label{6c}
\mathcal{W}=\int_{BZ}\frac{dk}{2\pi} \partial_{k}\phi_{k}
\end{equation}
where the phase $\phi_{k}$ is defined by $\tan\phi_{k}=\frac{h_{y}(k)}{h_{x}(k)}$. From the above formula, one gets $\mathcal{W}=1$ for $\Delta<0$ and zero otherwise. To illustrate the variation of winding number with $\Delta/t$, we plot  Fig. \ref{fig2}(b). The corresponding phase diagram is plotted in Fig. \ref{fig2}(c), which depicts the critical line $\Delta=0$ as the phase transition line. The region below and above this line indicates, respectively, a topologically nontrivial phase and a trivial phase.

\begin{figure}
   \vskip -.4 in
   \begin{picture}(100,100)
     \put(-30,0){
  \includegraphics[width=.6\linewidth]{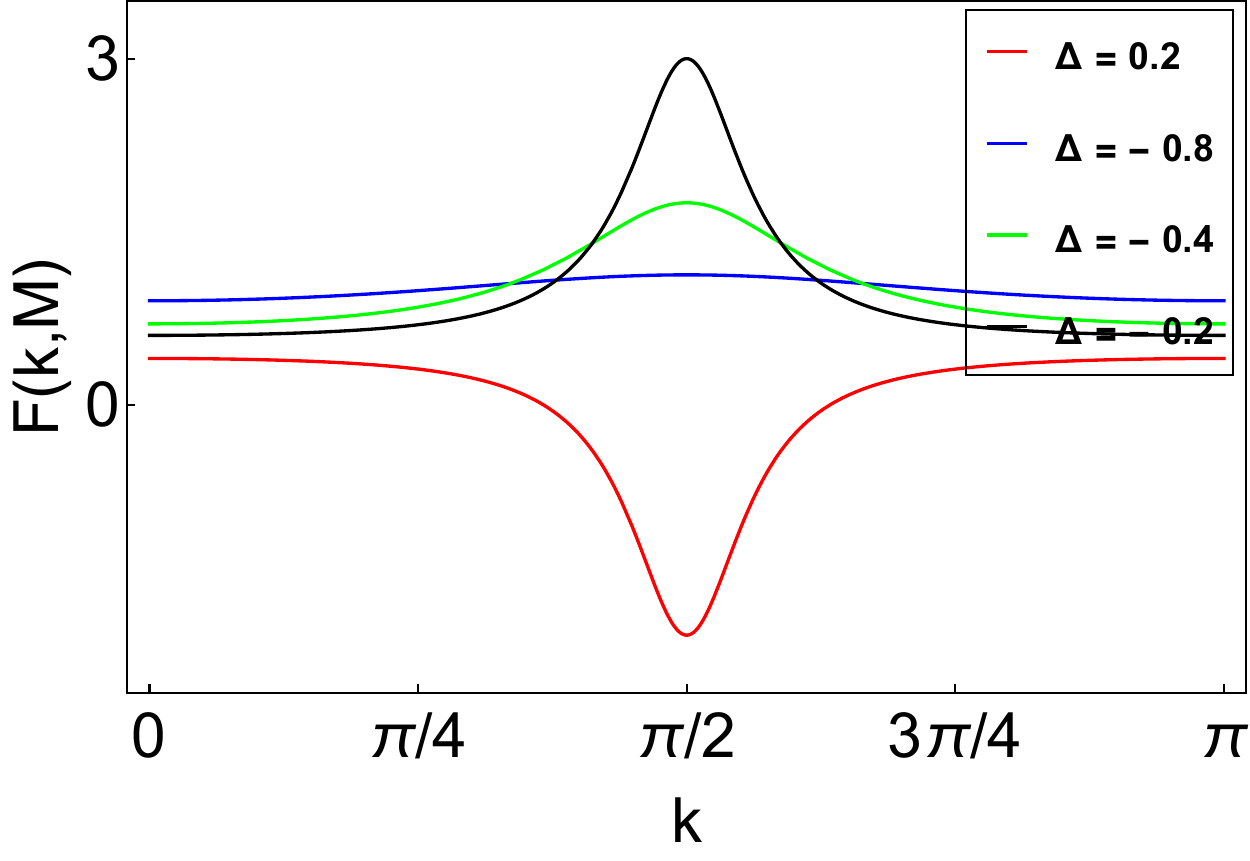}}
     \put(1,62){}
   \end{picture} 
\caption{The curvature function $F(k,\bf{M})$ near HSP $k_{0}=\pi/2$ for different values of $\Delta$. The divergence is prominent near the critical point $\Delta=0$. For the CRG procedure, $F(k_{0}+\delta k,\Delta)=F(k_{0},\Delta^{\prime})$, through which we notice the CRG flow $\Delta(=-0.2)\rightarrow\Delta^{\prime}(=-0.4)$, along which the divergence reduction occurs.} 
\label{fig1}
\end{figure}

To study the critical behavior near HSP, we first
determine the curvature function from Eq.(\ref{6c}) as

\begin{eqnarray}\label{7}
F(k,\bf{M})&=&\frac{d\phi_{k}}{dk}=\frac{d}{dk} \arctan\Big(\frac{Im[h(k)]}{Re[h(k)]}\Big)\nonumber\\
&=&\frac{d}{dk} \arctan\Big(\frac{(t-\Delta)\sin 2k}{(t+\Delta)+(t-\Delta)\cos 2k}\Big)\nonumber\\
&=&\frac{1}{2}-\frac{t \Delta}{t^2+\Delta^2+(t^2-\Delta^2)\cos 2k}
\end{eqnarray} 
 In order to verify that the curvature function manifests the Ornstein-Zernike form of Eq. (\ref{curvature2}), one can expand the function around the HSP, $k_{0}=\pi/2$. We plot $F(k,\bf{M})$ for different choices of $\Delta$ with fixed $t$ and at $k_{0}=\pi/2$ in Fig. \ref{fig1}. The plot clearly shows that the location of the critical point is at $\Delta=0$ and in that plot $F(k,\bf{M})$ develops a divergence at HSP $k_{0}=\pi/2$ as soon as the critical point is approached (for illustration purposes, one can compare the black line and green line) and crossing the critical point leads to the sign flipping of the divergence (notice the black and red line).

\subsubsection{Flow equations, fixed point, and critical point}
With the implication of the RG procedure discussed in Sec.\ref{sec2}, we get the following flow equations for the parameters $\bf{M}$=$(t,\Delta)$ near the high-symmetry point $k_{0}=\pi/2$ as
\begin{equation}\label{rg}
\frac{d\Delta}{dl}=\Delta\Big(\frac{t^2}{\Delta^2}-1\Big),
\end{equation}

\begin{equation}\label{rg1}
\frac{dt}{dl}=t\Big(1-\frac{t^2}{\Delta^2}\Big)
\end{equation}
The above two CRG Eqs. (\ref{rg}) and (\ref{rg1}) provides the flows of $(t,\Delta)$ as displayed in Fig. \ref{fig3}(a). We noticed that the CRG method applied at HSP $k_{0}=\pi/2$ exactly determines the critical lines associated with the gap closing HSP. Moreover, the divergence of the flow rate is visible at the horizontal line $\Delta=0$ which indicates the gap closing critical lines. This can be determined by considering the limit $\frac{d\Delta}{dl}\rightarrow \infty$ (the red line in Fig. \ref{fig3}(a) indicates the critical line). However, the fixed lines are estimated by using the limit $\frac{d\Delta}{dl}\rightarrow 0$ and calculating the fixed point at $\Delta=\pm t$. Among them, only the $\Delta=-t$ fixed point is topologically non-trivial and the trivial fixed point is at $\Delta=t$. These fixed points are shown by green lines in Fig. \ref{fig3}(a). The flow directs into the fixed lines making them stable fixed points. The scaling theory works well for $k_{0}=0$ though this is not a gap-closing point for $t\ne0$ (as discussed earlier). Hence we do not consider the scaling theory for $k_{0}=0$ here. The choice of different $k_{0}$, however, results in different speeds of convergence to the fixed point configuration\cite{rg}.

\subsubsection{Correlation length and critical exponents} 
As discussed in Sec.\ref{sec2}, one can observe the scaling bahavior of the curvature function, i.e., $F(k_{0},M)\sim|M-M_{c}|^{-\gamma}$ near the topological transition. By utilizing Eq. (\ref{curvature2}), the scaling form of correlation length can be estimated as $\xi_{k_{0}}\sim|M-M_{c}|^{-\nu}$. As mentioned in Sec. \ref{sec2}, $\nu$ and $\gamma$ represent critical exponents.

\begin{figure}
   \vskip -.4 in
   \begin{picture}(100,100)
     \put(-70,0){
  \includegraphics[width=.45\linewidth, height=1.25 in]{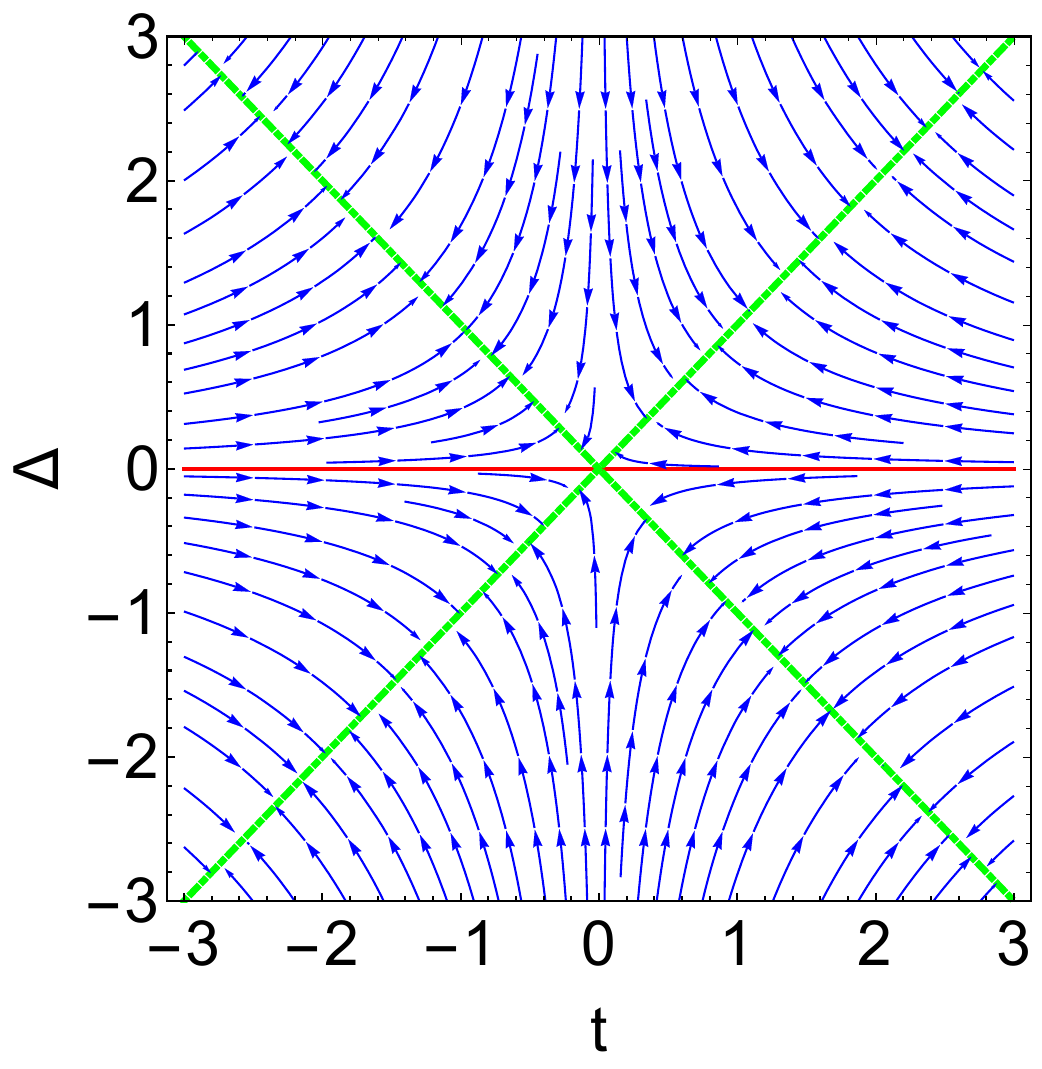}
  \includegraphics[width=.45\linewidth, height=1.25 in]{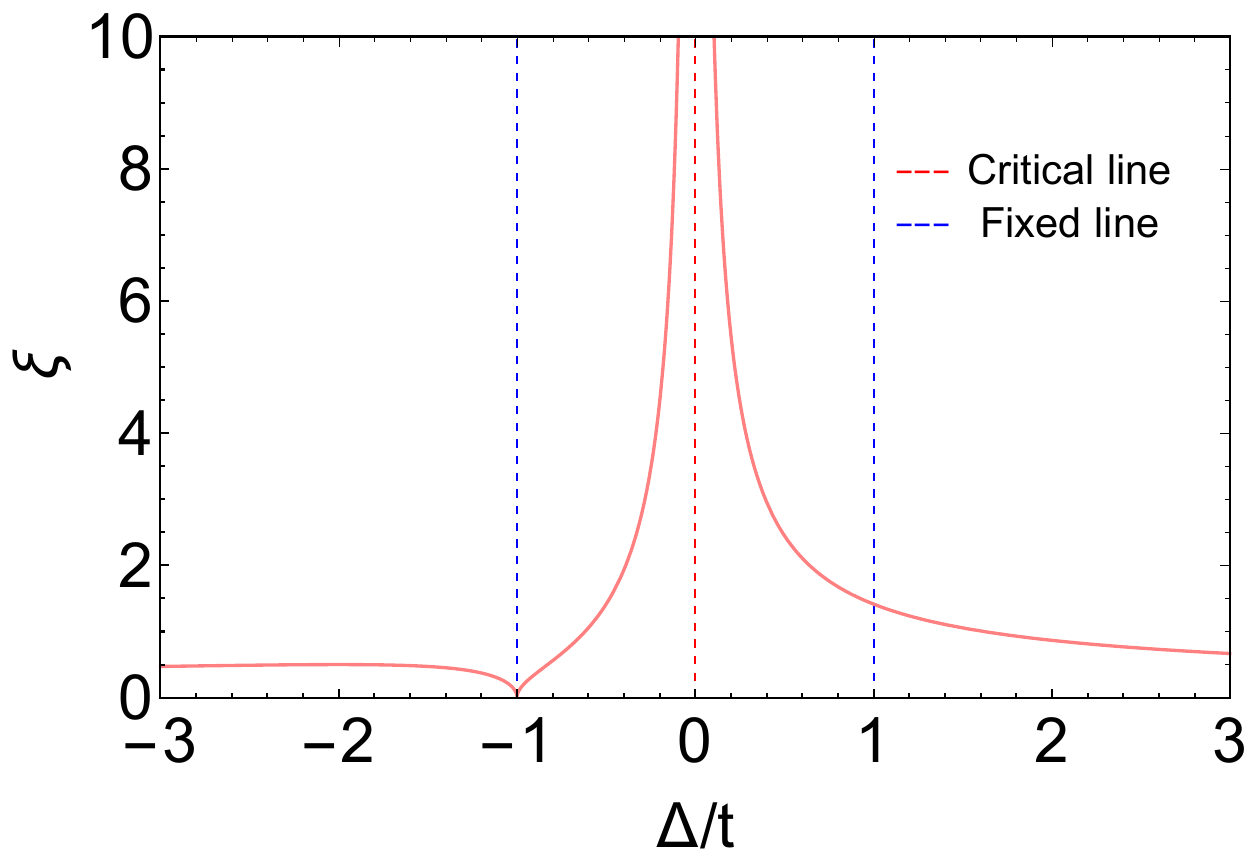}}
     \put(15,62){(a)}
     \put(125,58){(b)}
    \end{picture} 
  \vskip -0.1 in
\caption{(a) The RG flow near HSP, $k_{0}=\pi/2$. The red line indicates the critical points, while the green lines are for fixed points. (b) The correlation length as a function of $\Delta/t$.} 
\label{fig3}
\end{figure}

We can extract the exponents by enumerating the curvature function and Lorentzian expansion of it around HSP $k_{0}=\pi/2$. Now, the curvature function $F(k,\bf{M})$ calculated in Eq. (\ref{7}) at HSP becomes
\begin{equation}\label{co1}
F(k_{0},{\bf M})=\frac{1}{2}\Big(1-\frac{t}{\Delta}\Big)
\end{equation}
which diverges near the critical point, i.e., $\Delta_{c}=0$ as
\begin{equation}\label{co2}
F(k_{0},{\bf M})\sim-sgn\Big(\frac{t}{\Delta}\Big)|\Delta-\Delta_{c}|^{-1}
\end{equation}
Moreover, the Lorentzian form of $F(k,\bf{M})$, defined from Eq.(\ref{curvature2}), near $k_{0}=\pi$ becomes
\begin{equation}\label{co3}
\xi_{k_{0}}({\bf M})=\Big|\frac{t}{\Delta}\Big(1+\frac{t}{\Delta}\Big)\Big|^{1/2}\sim|\Delta-\Delta_{c}|^{-1}
\end{equation}
The behavior of this correlation function is depicted in Fig. \ref{fig3}(b), which shows the divergence of $\xi_{k_{0}}$ at the critical point $\Delta_{c}=0$ and vanishing nature at the non-trivial fixed point $\Delta=-t$ as expected\cite{rg}. Moreover, $\xi_{k_{0}}=1/\sqrt{2}$ for trivial fixed point $\Delta=t$. Comparing Eqs. (\ref{co2}) and (\ref{co3}), we immediately notice that the critical exponent defining the SSH chain TPT ($\mathcal{W}=1$) are $\gamma=\nu=1$\cite{edge1}, which is compatible with the scaling law. Recently, the same critical exponent was obtained for the $1D$ static Kiteav chain\cite{rg1}, making these models belong in the same universality class.

\subsubsection{Edge-state localization at the critical and fixed points}
 In $1D$, the end or edge states decay as
\begin{equation}\label{decay1}
\psi(n)\sim e^{-n/\xi_{edge}}
\end{equation}
where $\xi_{edge}$ represents the edge state decay length. However, note that this decay length of edge state is the same as the correlation length $\xi_{k_{0}}$\cite{edge1}. As $\xi_{k_{0}}$ diverges at $\Delta=0$ and vanishes at $\Delta=-t$, Eq. (\ref{decay1}) indicates that edge states exhibit complete delocalization at the critical point while maximum localization ($i.e.,$ decays sharply away from the edges) is observed at the non-trivial fixed point $\Delta=-t$ (see Appendix B for graphical illustration).

\subsection{Case II: $\theta=\frac{\pi}{2}$} 
\begin{figure}[!htbp]
    \centering
    \begin{tikzpicture}[scale=0.75] 

        \draw[thick, ->] (-4.2,0) -- (4.2,0) node[right] {$k$};
        
        \draw[dashed, gray] (-5.2,0) -- (-3,0);
        \draw[dashed, gray] (3,0) -- (5.2,0);

        \draw[ultra thick, blue] (-3,0) -- (3,0);

        \filldraw[black] (0,0) circle (2.5pt) node[above=4pt] {$\Gamma$};
        \node[below=6pt] at (0,0) {\small $0$};
        
        \filldraw[black] (3,0) circle (2.5pt) node[above=4pt] {$M$};
        \node[below=6pt] at (3,0) {\small $+\frac{\pi}{4}$};
        
        \filldraw[black] (-3,0) circle (2.5pt) node[above=4pt] {$M'$};
        \node[below=6pt] at (-3,0) {\small $-\frac{\pi}{4}$};

        \draw[latex-latex, thick, gray] (-3,-1.3) -- (3,-1.3) 
            node[midway, fill=white] {\small BZ Width $G = \frac{\pi}{2}$};
            
        \draw[dashed, gray] (-3,0) -- (-3,-1.4);
        \draw[dashed, gray] (3,0) -- (3,-1.4);

        \node[blue!80!black] at (1.5,0.4) {\scriptsize Non-HSP};
        \node[blue!80!black] at (-1.5,0.4) {\scriptsize Non-HSP};

    \end{tikzpicture}
    \caption{Geometric representation of the reduced $1D$ FBZ for the $n=4$ modulated lattice configuration. The high symmetry points (HSPs) are located at $\Gamma$ and the boundaries $M, M'$, while the interior domains consist entirely of non-high symmetry points (Non-HSPs).}
    \label{fig:reduced_bz}
\end{figure}
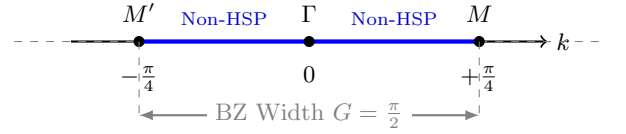

For this cases, the unit cell contains four sublattices and the BZ is folded due to the increased sublattices. So the size of the FBZ boundary reduces $k\in [-\pi/4,~\pi/4]$ with reduced reciprocal vector being $G=\pi/2$ and within the reduced FBZ the center is at $k=0~(\Gamma)$ and boundary is at $k=\pi/4~(M)$ and $-\pi/4~(M^{\prime})$. Thus, we now get two HSP at $k_{0}=0$ and $k_{0}=\pm\pi/4$ which is shown schematically in Fig. \ref{fig:reduced_bz}.

\subsubsection{Hamiltonian and phase diagram}
The Bloch Hamiltonian in $k$-space takes the form of:
\begin{equation}\label{18aa}
H_{k} = 
\begin{pmatrix}
0 & (t+\Delta) & 0 & t e^{-4ik} \\
(t+\Delta) & 0 & t & 0 \\
0 & t & 0 & (t-\Delta) \\
t e^{4ik} & 0 & (t-\Delta) & 0
\end{pmatrix}.
\end{equation}
The energy eigenvalues for the above Hamiltonian can be estimated as
\begin{equation}\label{19}
E(k)_{\pm}=\pm\sqrt{2t^2+\Delta^2\pm t\sqrt{2t^2+6\Delta^2+2(t^2-\Delta^2) \cos4k}}
\end{equation}
It produces four energy bands in the spectrum. The energy gap closes for the momenta given by

\begin{equation}\label{19a}
k=k_0=\pm\frac{1}{4}\arccos\Big[\frac{2t^4-2t^2\Delta^2+\Delta^4}{2t^2(t^2-\Delta^2)}\Big]
\end{equation}

\begin{figure}
   \vskip -.4 in
   \begin{picture}(100,100)
     \put(-70,0){
  \includegraphics[width=.45\linewidth, height=1.25 in]{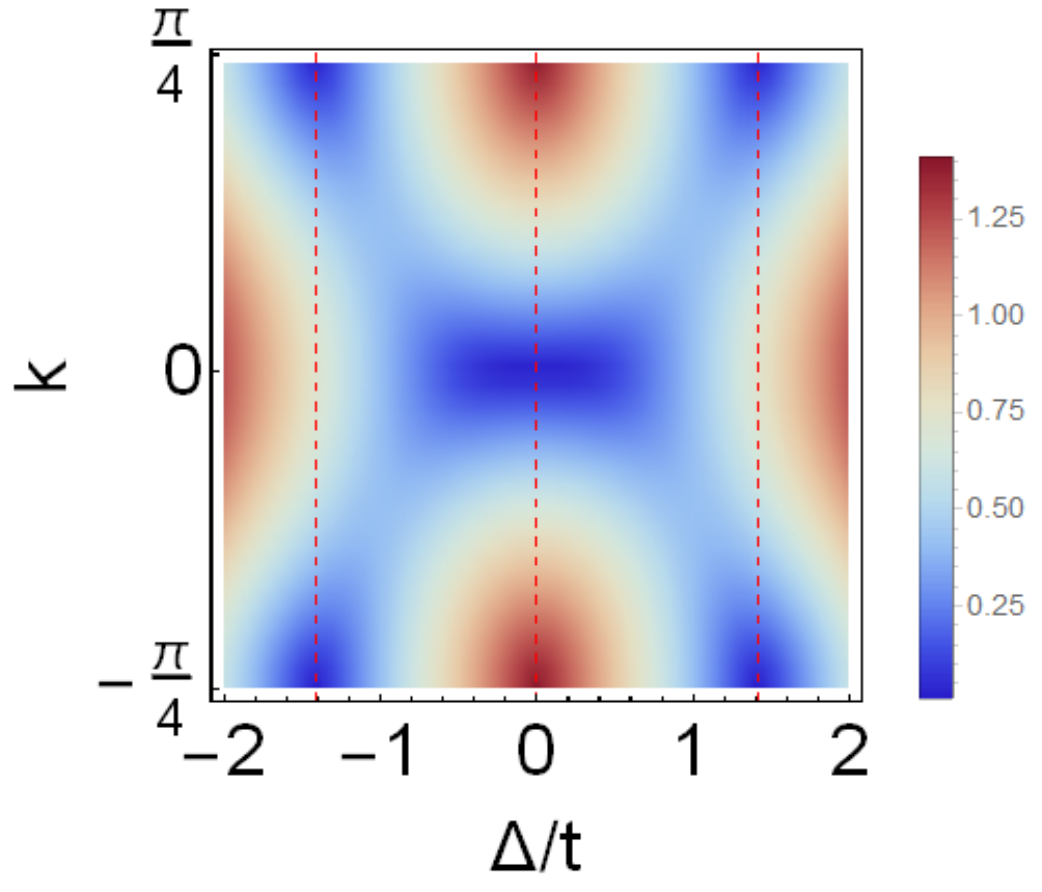}
  \includegraphics[width=.45\linewidth, height=1.25 in]{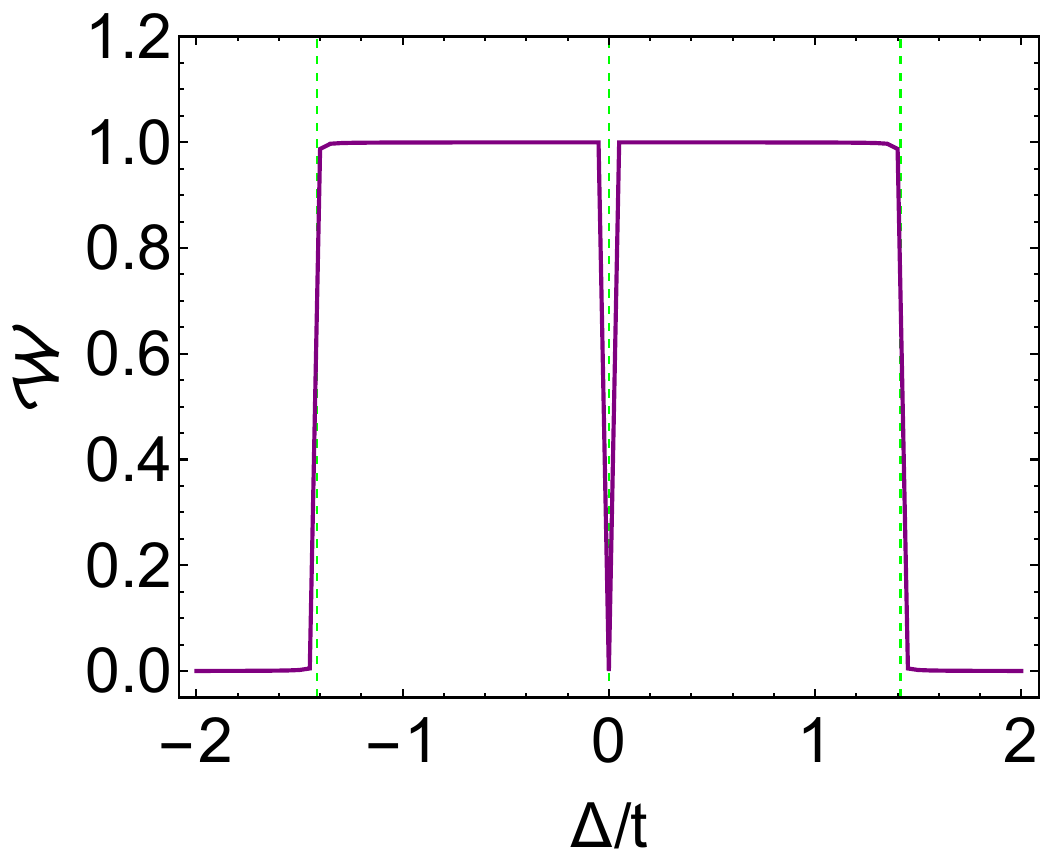}}
     \put(-10,62){(a)}
     \put(125,58){(b)}
    \end{picture}\\
     \vskip -.00004 in
   \begin{picture}(100,100)
     \put(-10,0){
       \includegraphics[width=.48\linewidth,height=1.25 in]{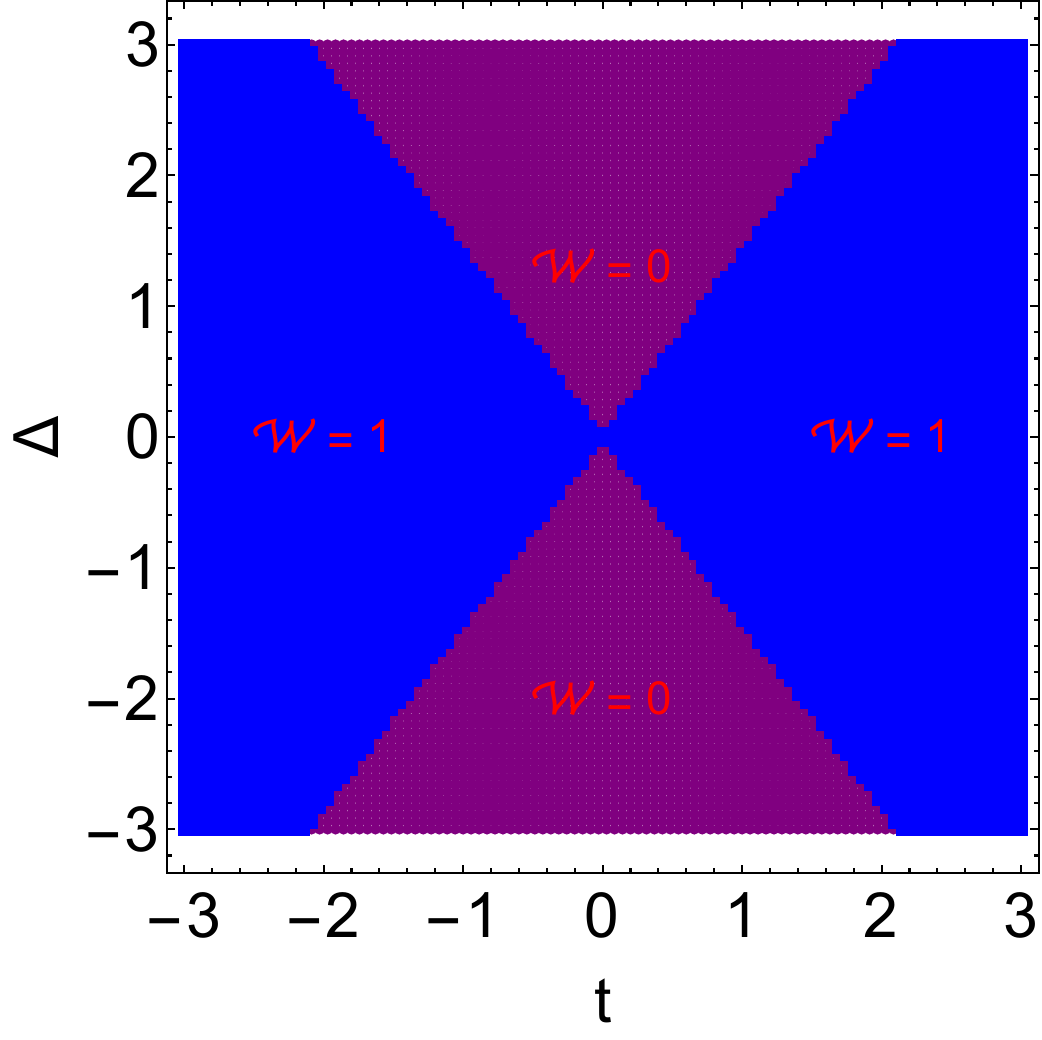}}
   \put(-10,58){(c)}
   \end{picture}
  \vskip -0.1 in
\caption{(a) The bulk gap in the $(\Delta/t-k)$-plane for four-site SSH chain. Here, $|\Delta|=\sqrt{2}t$ and $\Delta=0$ are the gap closing line when $k=\pm\pi/4$ and $k=0$ respectively. The deep blue in the color bar indicates zero bulk gap. (b) The variation of winding number with $\Delta/t$. (c) The phase diagram showing the topological (blue region) and trivial (purple) region. The red and green vertical dashed lines in (a) and (b) refer to the
position of $\Delta = 0$ and $|\Delta/t| = \sqrt{2}$.} 
\label{figpiby2}
\end{figure}

For the system under PBC, the above relation gives the occurrence of a gap-closing point in the energy spectrum at $\Delta=0$ for HSP $k=0$ as well as at $|\Delta/t|=\sqrt{2}$ for HSP $k_0=\pm\pi/4$ (see Fig. \ref{figpiby2}(a)). The lowest positive energy band gap at $k_{0}=0$ for $\Delta\ne0$ becomes $E_{gap}\sim|\Delta|^{2}$ and it would be $E_{gap}\sim |\sqrt{2}t\pm \Delta|$ at the boundary of the FBZ ($i.e.,$ for $k\in [-\pi/4,~\pi/4]$) when $|\Delta/t|\ne\sqrt{2}$. These result in quadratic gap closing exactly at $\Delta=0$ point for the system (both under PBC and OBC) but a linear (Dirac type) band touching (both under PBC and OBC) at $|\Delta/t|=\sqrt{2}$ (see Ref.\cite{mandal} and Appendix A for graphical illustration). Importantly, the winding number, at the $\Delta=0$ point, doesn’t change (trivial-to-trivial), indicating a trivial metallic point due to a second-order topological transition (because the modulation parameter $\Delta$ enters quadratically in the effective mass or $E_{gap}$). However, the winding number changes on either side of $|\Delta/t|=\sqrt{2}$ (notice Fig. \ref{figpiby2}(b)) and is associated with the gap-closing topological phase transition point\cite{mandal}, which is a conventional first-order (Dirac-type) topological transition.  Thus, TPT occur only at HSP $k_{0}=\pm\pi/4$ while HSP $k_{0}=0$ doesn't contribute to the system's topology. The chain under OBC, however, produces two ZES and two nonzero energy in-gap states\cite{mandal}.

The Hamiltonian $H_{k}$ in Eq. (\ref{18aa}) respects time reversal symmetry $TH(k)T^{-1}=H(-k)$ with $T$ denotes complex conjugation\cite{symmetry,symmetry1}. It also has chiral symmetry $SH(k)S^{-1}=-H(k)$ follows from the particle-hole symmetry ($C$). The chiral-symmetric operator is given by $\hat{S}=\Gamma_{4}=I_{2}\otimes\sigma_{z}$ with $I_{2}$ and $\sigma_{z}$ represents the $2\times2$ identity matrix and third component of Pauli matrices respectively\cite{mandal}. Thus, the model falls in the $BDI$ class in the tenfold way classification scheme. However, by employing the unitary transformation with the unitary matrix 
\begin{equation}\label{8}
U=
\begin{pmatrix}
1 & 0 & 0 & 0 \\\
0 & 0 & 1 & 0\\\
0 & 1 & 0 & 0\\\
0 & 0 & 0 & 1
\end{pmatrix}
\end{equation}
convert the Hamiltonian $H_{k}$ in Eq. (\ref{18aa}) into the block off-diagonal form $H_{k}\rightarrow UH(k)U^{-1}=\begin{pmatrix}
0 & V \\
V^{\dagger} & 0 
\end{pmatrix}$ with the upper off-diagonal block read as
\begin{equation}\label{7a}
V(k)=
\begin{pmatrix}
 (t+\Delta) & te^{-4ik} \\
 t & (t-\Delta)
\end{pmatrix}
\end{equation}
This model, with four bands, exhibits non-trivial topological phases\cite{mandal} characterized by the winding number (the formula can also hold for the model having bands more than four) given by\cite{s1,s2,s1a,s3,s3a}
\begin{equation}\label{9}
\mathcal{W}=\int_{BZ}\frac{dk}{2\pi i} \partial_{k}[\ln {\bf Det}[V(k)]]
\end{equation}
with winding vector ${\bf Det}[V(k)]=(t^2-\Delta^2)-t^2e^{-4ik}=R(k)e^{i \phi(k)}$ (where $\phi_{k}$ is the phase of ${\bf Det}[V(k)]$). By examining $\mathcal{W}$ over the parameter space parameterized by $\mathcal{M}=(\Delta,t)$, one attains the non-trivial topological phases (NTP) with ${\mathcal W}=1$ till $0<|\Delta/t|<\sqrt{2}$ and trivial topological phase (TTP) with ${\mathcal W}=0$ for $|\Delta/t|>\sqrt{2}$ or $\Delta=0$. Therefore, the TPT is visible at the critical line $\Delta=\pm\sqrt{2}t$ for this case, as plotted in Fig. \ref{figpiby2}(b). Fig. \ref{figpiby2}(b) also shows a trivial-to-trivial phase transition at $\Delta=0$ point, resulting in ${\mathcal W}=0$. The corresponding phase diagram is illustrated in Fig. \ref{figpiby2}(c). Importantly, the mere sign change in winding number (i.e., $\mathcal{W}\rightarrow-\mathcal{W}$) is noticeable as soon as we proceed by considering the lower off-diagonal block $V^{\dagger}(k)$. The sign changes of winding number now enforce the chiral symmetry operator to become $\hat{S}=-\Gamma_{4}$\cite{mandal}.

In order to illustrate the critical behavior near the TPT, we eliminate the curvature function from Eq. (\ref{9}) for this four-band model as:

\begin{eqnarray}\label{11}
F(k,\bf{M})&=&\frac{d\phi_{k}}{dk}=\frac{d}{dk} \arctan\Big(\frac{Im[{\bf Det}[V(k)]]}{Re[{\bf Det}[V(k)]]}\Big)\nonumber\\
&=&\frac{d}{dk} \arctan\Big(\frac{-t^2\sin k}{(t^2-\Delta^2)-t^2\cos k}\Big)\nonumber\\
&=&\frac{4}{2+\frac{\Delta^4-2t^2\Delta^2}{t^4+(\Delta^2-t^2)\cos4k}}
\end{eqnarray} 

\begin{figure}
   \vskip -.4 in
   \begin{picture}(100,100)
     \put(-70,0){
  \includegraphics[width=.45\linewidth]{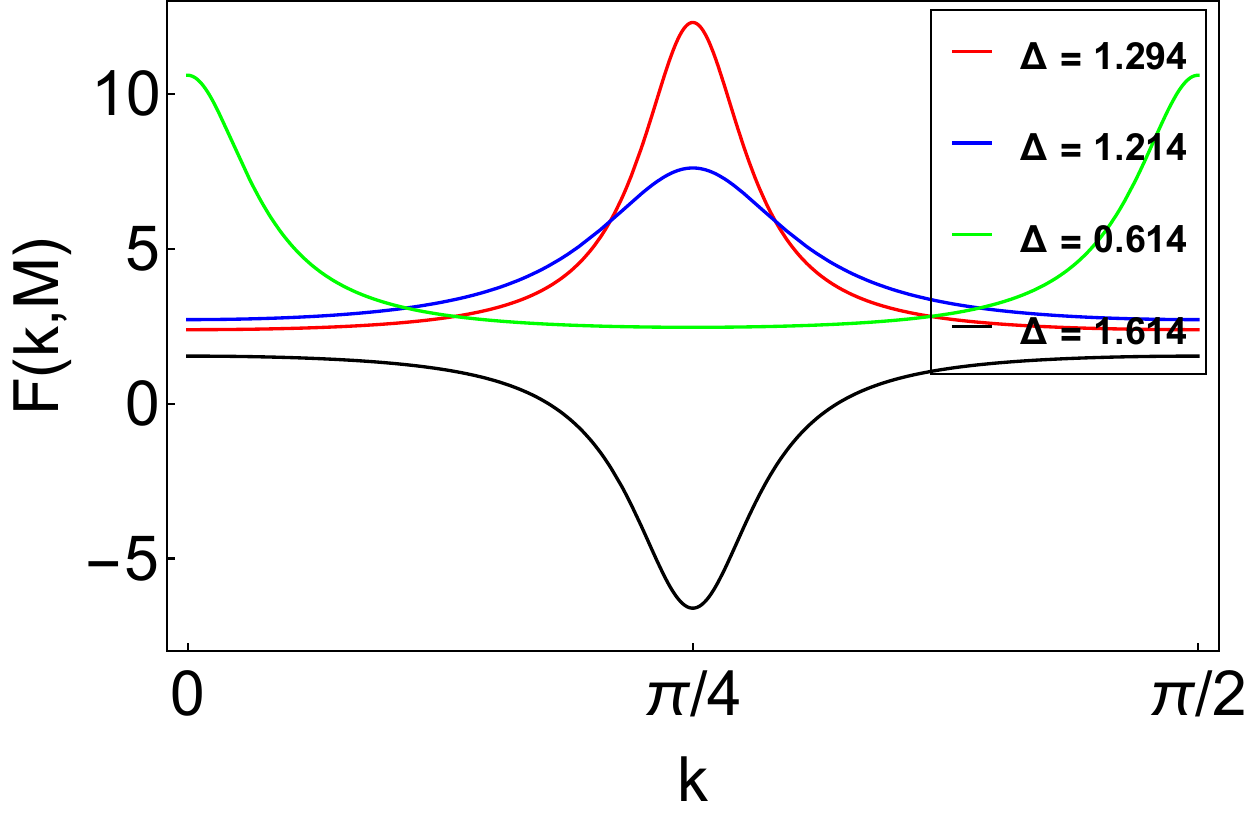}
   \includegraphics[width=.45\linewidth]{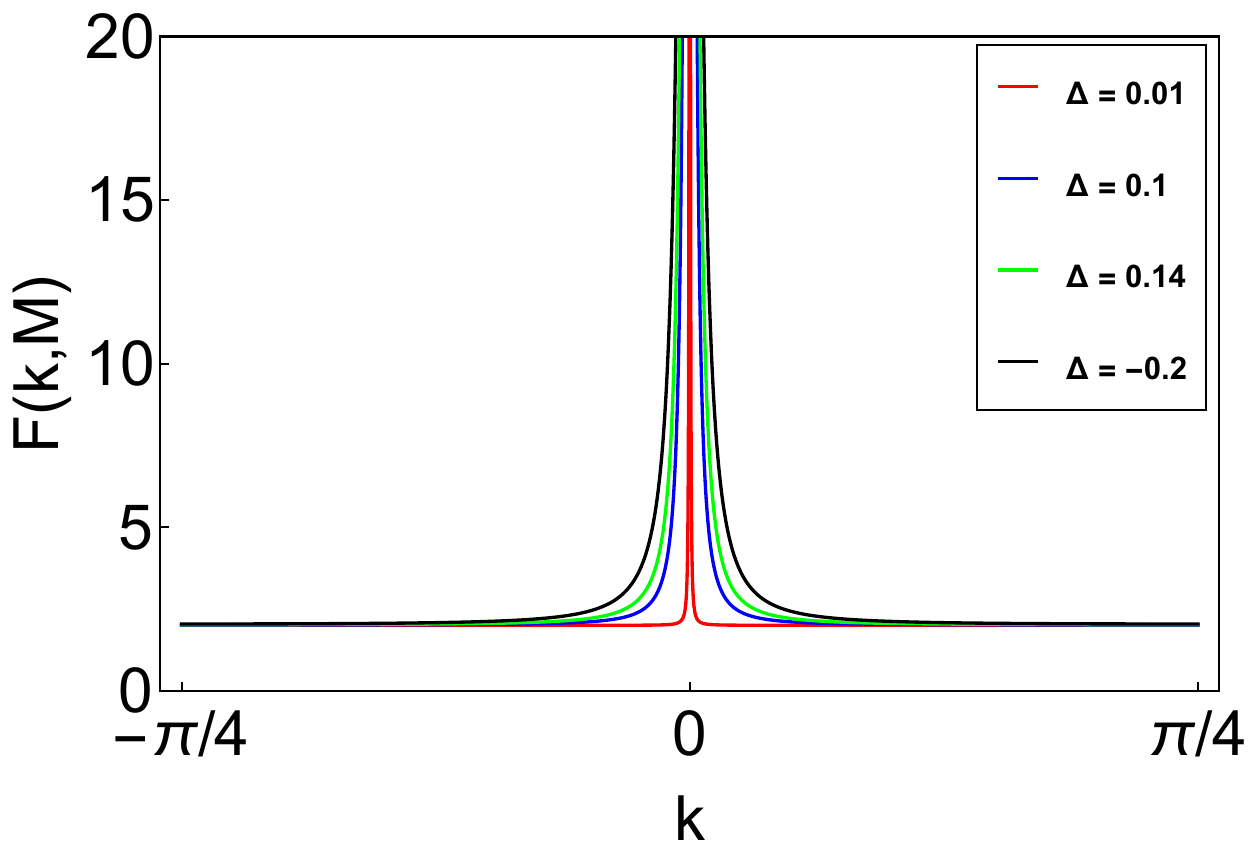}}
     \put(-10,62){(a)}
      \put(125,62){(b)}
   \end{picture} 
\caption{(a) The behaviour of curvature function $F(k,\bf{M})$ around HSP $k_{0}=\pi/4$ for different choices of $\Delta$. The divergence is prominent near the critical point $|\Delta/t|=\sqrt{2}$. The divergence reduction occurs through $\Delta~(=1.294)\rightarrow\Delta^{\prime}~ (=1.214)$ following the CRG procedure, $F(k_{0}+\delta k,\Delta)=F(k_{0},\Delta^{\prime})$. (b) Faster divergence of $F(k,\bf{M})$ near non-topological HSP $\Delta=0$ around $k_{0}=0$.} 
\label{fig4}
\end{figure}

Now, we plot Fig. \ref{fig4} to demonstrate the divergence of the curvature function and its sign flip across the gap-closing HSP. Unlike the previous case, here the divergence of $F(k,{ \bf M})$ occurs at $k_{0}=\pi/4$ when we move toward the critical point $|\Delta/t|=\sqrt{2}$. Here, the divergence reduction occurs through the CRG flow $\Delta~(=1.294)\rightarrow\Delta^{\prime}~ (=1.214)$ (notice Fig. \ref{fig4}(a)). Additionally, this case also shows much faster divergence of $F(k,{ \bf M})$ at the non-topological gapless point $\Delta=0$ for $k_{0}=0$ as compared to the topological transition point $|\Delta/t|=\sqrt{2}$ as shown in Fig. \ref{fig4}(b).
\subsubsection{Flow equations, fixed points and critical points}
Again, we apply the RG procedure discussed in Sec.\ref{sec2} and attain the following flow equations in the parameters $\bf{M}$=~$(t,\Delta)$ near the high-symmetry point $k_{0}=\pi/4$ as
\begin{equation}\label{rg2}
\frac{d\Delta}{dl}=\frac{4\Delta(\Delta^2-t^2)}{\Delta^2-2t^2},
\end{equation}

\begin{equation}\label{rg3}
\frac{dt}{dl}=-\frac{4t(\Delta^2-t^2)}{\Delta^2-2t^2}
\end{equation}

Again, the RG flow equations near $k_{0}=0$ is calculated as
\begin{equation}\label{rg4}
\frac{d\Delta}{dl}=\frac{4(\Delta^4-3\Delta^2t^2+2t^4)}{\Delta^3},
\end{equation}

\begin{equation}\label{rg5}
\frac{dt}{dl}=-\frac{2t(\Delta^4-3\Delta^2t^2+2t^4)}{\Delta^4}
\end{equation}

\begin{figure}
   \vskip -.4 in
   \begin{picture}(100,100)
     \put(-70,0){
  \includegraphics[width=.45\linewidth, height=1.25 in]{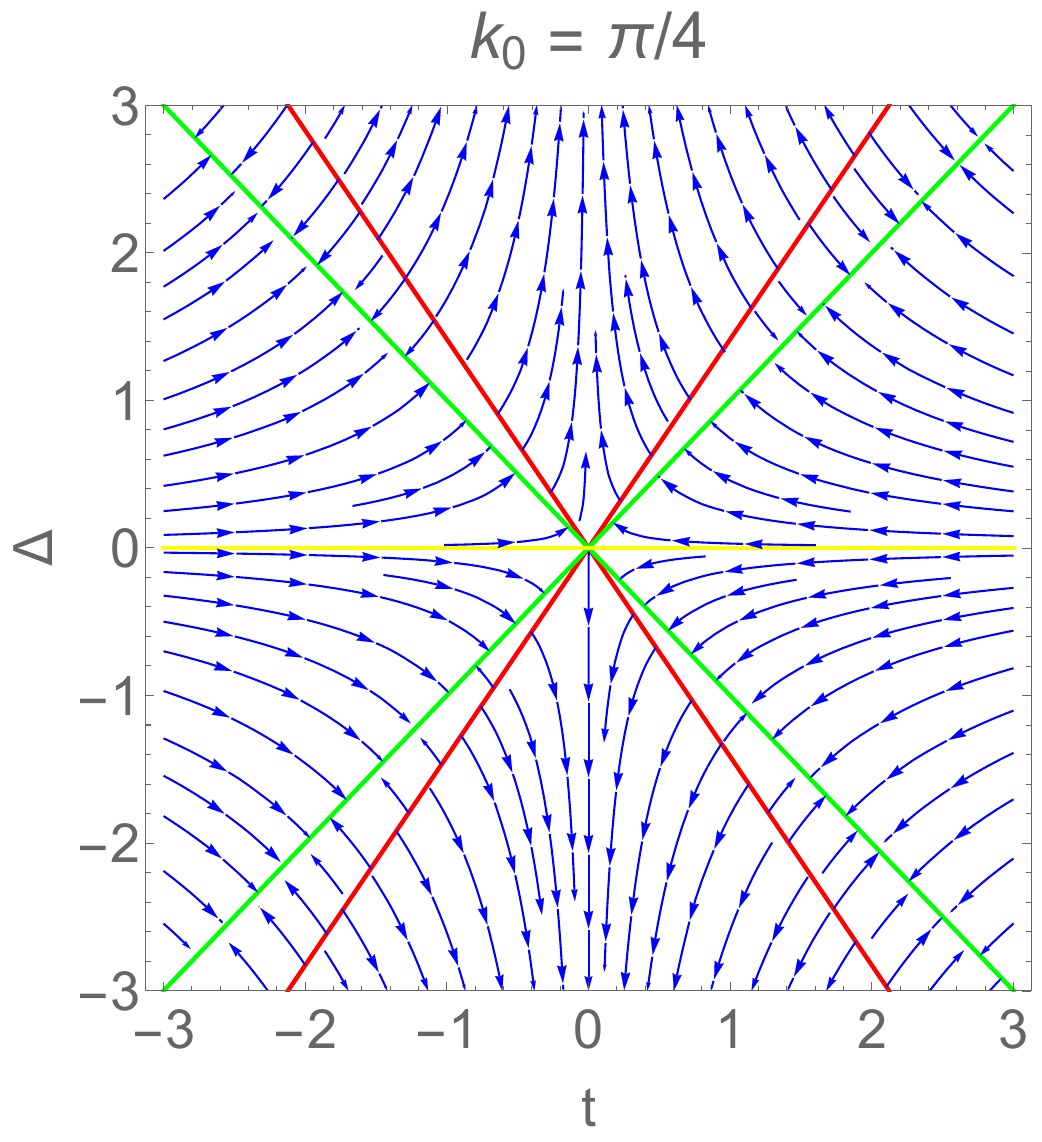}
  \includegraphics[width=.45\linewidth, height=1.25 in]{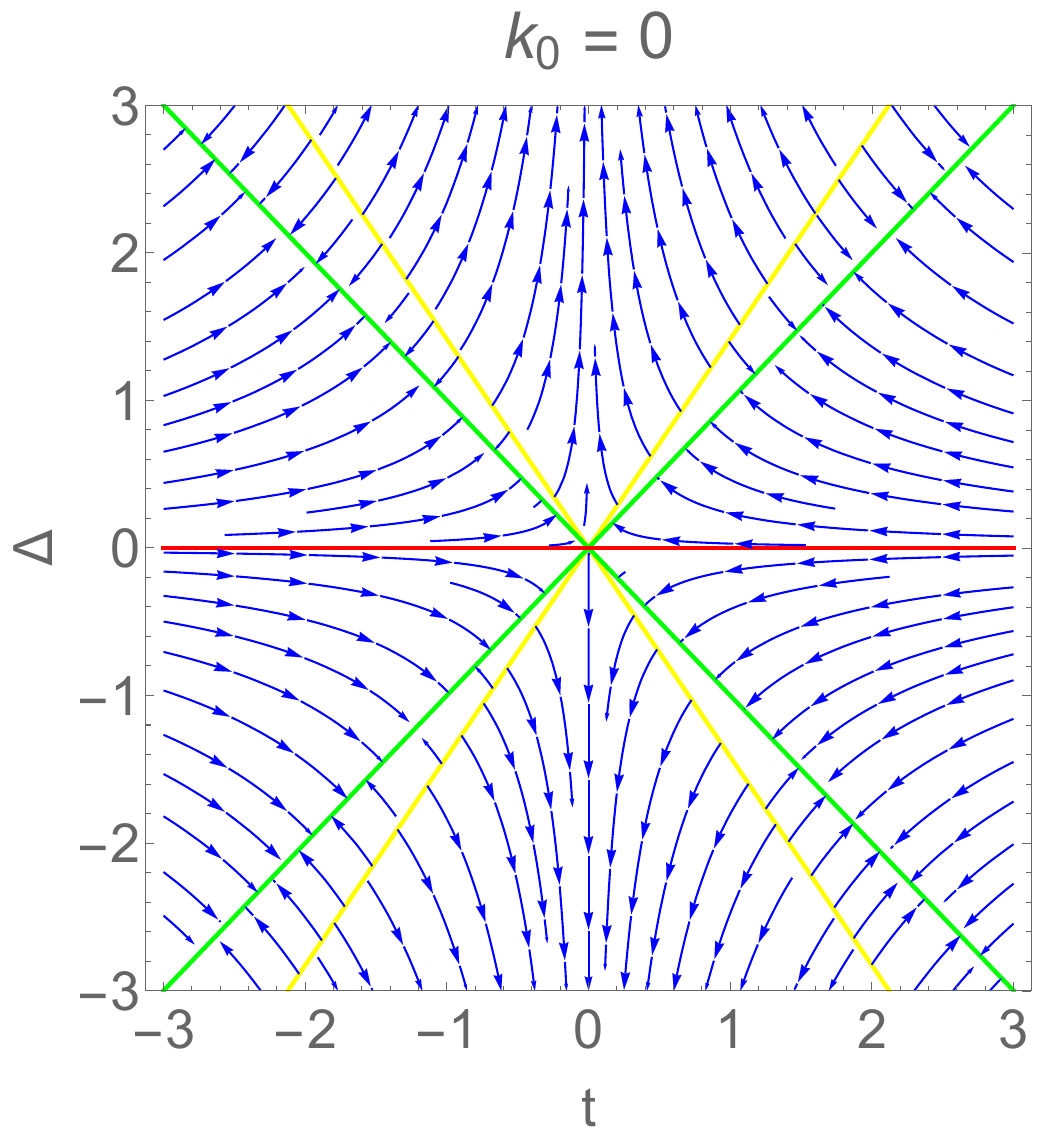}}
     \put(10,62){(a)}
     \put(125,58){(b)}
    \end{picture} 
  \vskip -0.1 in
\caption{ The RG flow diagram near HSP $k_{0}=\pi/4$ (a) and $k_{0}=0$ (b). The red line in (a) indicates the topological critical points, while the same in (b) indicates the non-topological metallic-like point. The green lines and yellow lines in (a) and (b) are for fixed points.} 
\label{fig3a}
\end{figure}

The RG flows of ($t,\Delta$) for this case are represented by the Eqs. (\ref{rg2})-(\ref{rg5}). It is noticed from the above Eqs. (\ref{rg2}) and (\ref{rg3}), that the right-hand side of the RG equations in the infinite limit correctly estimates the critical point at $|\Delta|=\sqrt{2}t$ for Dirac-like HSP $k_{0}=\pi/4$ and the vanishing limit of these equations gives a fixed line at $\Delta=0,~\pm t$. Moreover, the infinite limit of Eqs. (\ref{rg4}) and (\ref{rg5}) predicts the second-order trivial transition point at $\Delta=0$ while the vanishing limit gives the fixed points at $\Delta=\pm t, ~\pm \sqrt{2}t$ for metallic-like HSP $k_{0}=0$. Here, the fixed points at $\Delta=\pm t$ are topologically non-trivial. The RG flow diagram for HSPs $k_{0}=\pi/4$ and $k_{0}=0$ are reported in Figs. \ref{fig3a}(a) and (b) respectively. We notice from the flow diagram that CRG implemented at $k_{0}=\pi/4$ and $k_{0}=0$ correctly regenerates the phase diagram as shown in Fig. \ref{figpiby2}(c). The divergence of flow rate discernible along the line $\Delta=\pm \sqrt{2}t$ and $\Delta=0$, which are shown by a red line in Figs. \ref{fig3a}(a) and (b) respectively. 
The fixed lines for both cases are shown in green and yellow color where for the green lines, the flow rate converges with $\frac{d\Delta}{dl},\frac{dt}{dl}\rightarrow 0$, signifying stable fixed points. Conversely, we notice the flow rate for yellow lines diverges with $\frac{d\Delta}{dl},\frac{dt}{dl}\rightarrow 0$, indicating unstable fixed points.

By emphasizing the RG flow diagram Fig. \ref{fig3a}(a), we clearly see that it not only reproduces the critical line $|\Delta|=\sqrt{2}t$, but the line passing through the unstable fixed points also regenerates the line where the gap vanishes at the HSP $k_{0}=0$. Similarly, the flow diagram Fig. \ref{fig3a}(b), aside from reproducing the $\Delta=0$ second-order trivial transition line, the unstable fixed point line also reproduces the $|\Delta|=\sqrt{2}t$ critical line almost completely.

\subsubsection{Correlation length and critical exponents}
The enumeration of the curvature function and Lorentzian expansion of it around HSP helps us to extract the critical exponents. Now, the curvature function $F(k,\bf{M})$ calculated in Eq. (\ref{11}) takes the following form:
\begin{equation}\label{co1a}
F(k_{0},{\bf M})=\begin{cases}
{-\frac{4t^2}{\Delta^2-2t^2}}, & \text{at HSP $k_{0}=\pi/4$}\\
~~~~~~\frac{4t^2}{\Delta^2}, & \text{at HSP $k_{0}=0$}
\end{cases}
\end{equation}

It diverges near the criticality, i.e., $\Delta_{c}=\sqrt{2}t$, as
\begin{equation}\label{co2a}
F(k_{0},{\bf M})\sim-sgn(\Delta^2-2t^2)|\Delta^2-\Delta_{c}^2|^{-1}
\end{equation}
and the correlation length $\xi_{k_{0}}$, close to the criticality, becomes
\begin{equation}\label{co3a}
\xi_{k_{0}}({\bf M})=\sqrt{8}\frac{\sqrt{\Delta^2(\Delta^2-t^2)}}{|\Delta^2-2t^2|}\sim |\Delta^2-\Delta_{c}^2|^{-1}
\end{equation}

Now, the curvature function, following Eq. (\ref{co1a}), around the quadratic gap closing point $\Delta_{c_{1}^{\prime}}=0$ is approximated as $F(k,\bf{M})$ $\sim |\Delta-\Delta_{c_{1}^{\prime}}|^{-2}$ and $\xi_{k_{0}}$ near this point is estimated as
\begin{equation}\label{co3b}
\xi_{k_{0}}({\bf M})=\frac{2t}{|\Delta|^2}\sqrt{4t^2-6\Delta^2};
\end{equation}
which in the limit $\Delta^2/t^2<<1$ reads $\xi_{k_{0}}=\frac{4t^2}{|\Delta|^2}\sim |\Delta-\Delta_{c_{1}^{\prime}}|^{-2}$.

\begin{figure}
   \vskip -.4 in
   \begin{picture}(100,100)
     \put(-70,0){
  \includegraphics[width=.45\linewidth, height=1.25 in]{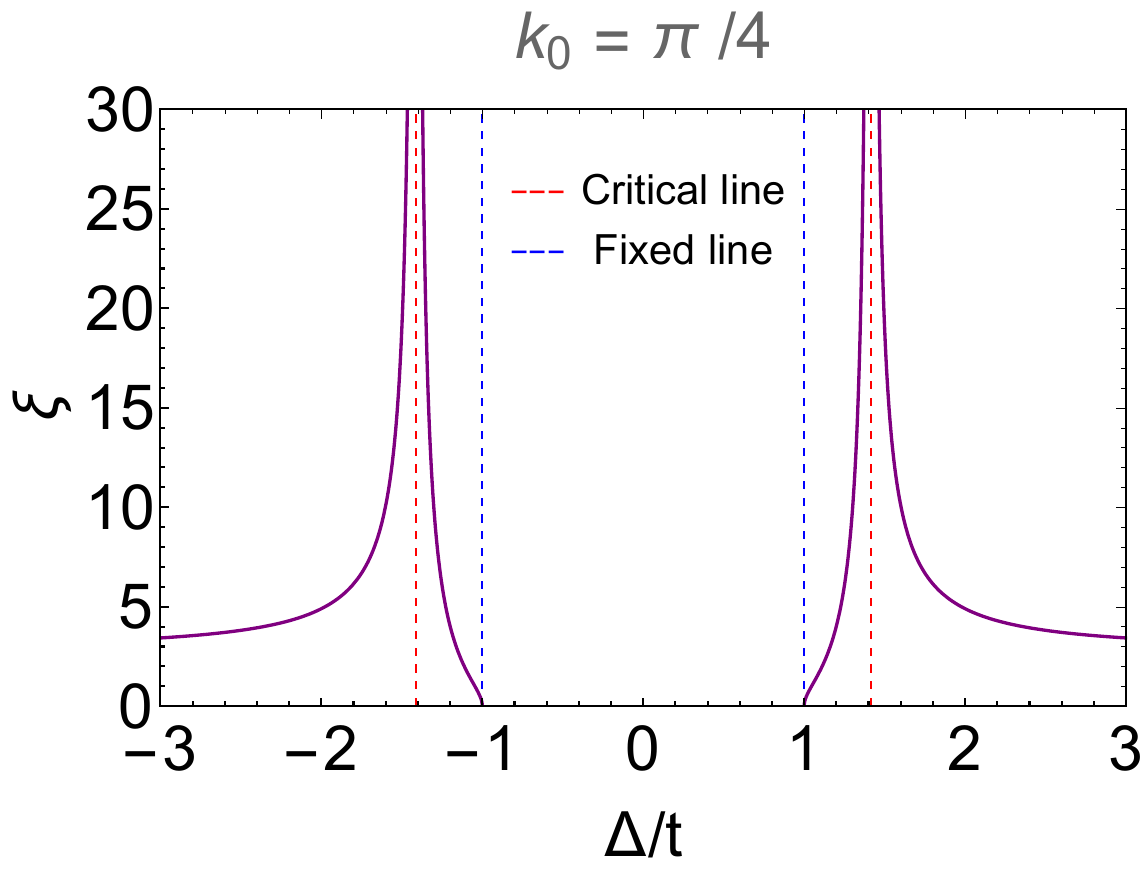}
  \includegraphics[width=.45\linewidth, height=1.25 in]{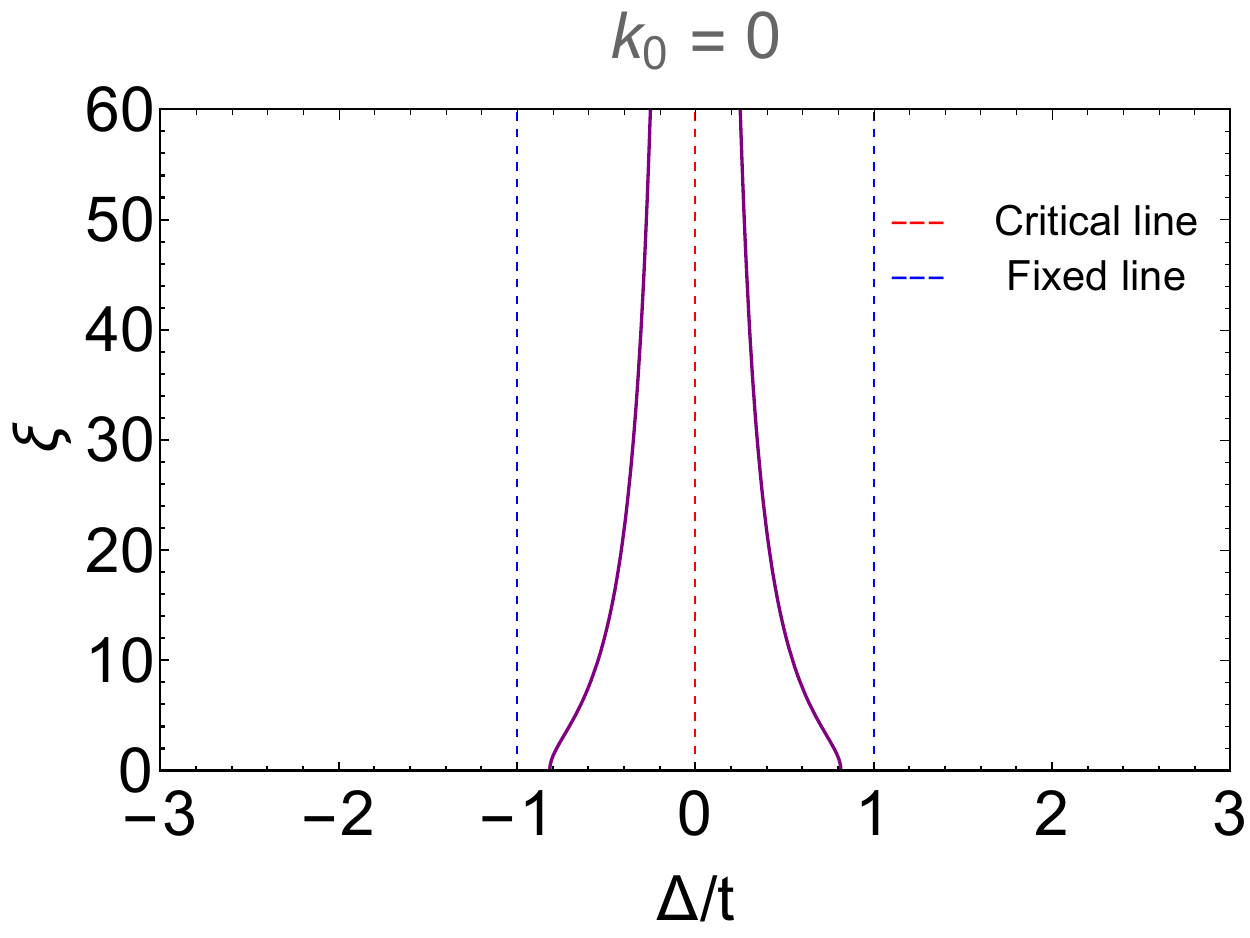}}
     \put(-15,62){(a)}
     \put(75,60){(b)}
    \end{picture} 
  \vskip -0.1 in
\caption{ The correlation length as a function of $\Delta/t$ for (a) $k_{0}=\pi/4$ and (b) $k_{0}=0$.} 
\label{fig3b}
\end{figure}

The behavior of the correlation function for $k_{0}=\pi/4$ and $k_{0}=0$ are depicted in Fig. \ref{fig3b}.  In Fig. \ref{fig3b}(a), we notice $\xi_{k_{0}}$ diverges at the critical point $\Delta_{c}=\pm\sqrt{2}t$ while it vanishes at the non-trivial fixed point $\Delta=\pm t$ as expected. The critical exponent for $k_{0}=\pi/4$ becomes $\gamma=\nu=1$ as can be found from Eqs. (\ref{co2a}) and (\ref{co3a}). Interestingly, as soon as $\Delta$ approaches zero, the gap vanishes and $\xi_{k_{0}}$ diverges as $|\Delta-\Delta_{c_{1}^{\prime}}|^{-2}$, indicating a second order-phase transition which is shown in Fig. \ref{fig3b}(b). Thus, the critical exponent for $k_{0}=0$ comes out to be $\gamma=\nu=2$. It is noticed from Fig. \ref{fig3b}(b) that $\xi_{k_{0}}$ will show much faster divergence approaching $\Delta=0$.

In brief, the curvature function diverges more sharply around the metallic-like point $\Delta=0$, giving $\gamma=\nu=2$. On the other hand, one can notice comparatively slower divergence near the critical line $|\Delta|=\sqrt{2}t$, which gives $\gamma=\nu=1$. Thus, the same model for two different points, i.e., for $\Delta=0$ and $|\Delta/t|=\sqrt{2}$ belong to different universality classes due to different critical exponents. It is noteworthy to mention here that the critical exponents for the model considering $\Delta=0$ come out to be $\gamma=\nu=2$, which is different from the known critical exponents calculated for the other models\cite{rg1,edge1}, making our result more interesting. Here, we may conclude that the different kinds of gap-closing nature of these points give rise to different critical exponents.

\subsubsection{Edge-state localization at the critical and fixed points}
We should mention here that the strength of the divergence of correlation length also determines how quickly edge-states penetrate into the bulk near and exactly at the position of critical, metallic-like and fixed points.
Here, it is evident that the edge modes become completely delocalized at the metallic-like and critical points, namely at $|\Delta| = 0$ and $\sqrt{2}t$. As the system moves away from these points towards the stable fixed points $\Delta = \pm t$, the edge modes become progressively more localized. 
Interestingly, the localization is stronger near $|\Delta/t| = \sqrt{2}$ than near $\Delta = 0$, as reflected by the faster decay of the edge-state amplitudes into the bulk (this is because of the faster divergence of $\xi$ around $\Delta = 0$ compared to $|\Delta/t| = \sqrt{2}$). Moreover, the edge modes exhibit their maximum degree of localization at the stable fixed points
$\Delta = \pm t$. We studied the graphical nature of the same in
Appendix B.

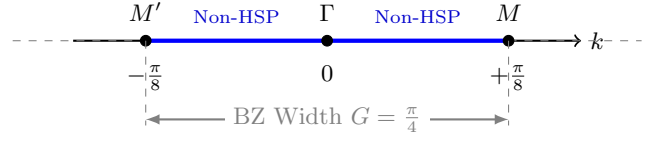
\begin{figure}[!htbp]
    \centering
    \begin{tikzpicture}[scale=0.8]

        \draw[thick, ->] (-4.2,0) -- (4.2,0) node[right] {$k$};
        
        \draw[dashed, gray] (-5.2,0) -- (-3,0);
        \draw[dashed, gray] (3,0) -- (5.2,0);

        \draw[ultra thick, blue] (-3,0) -- (3,0);

        \filldraw[black] (0,0) circle (2.5pt) node[above=4pt] {$\Gamma$};
        \node[below=6pt] at (0,0) {\small $0$};
        
        \filldraw[black] (3,0) circle (2.5pt) node[above=4pt] {$M$};
        \node[below=6pt] at (3,0) {\small $+\frac{\pi}{8}$};
        
        \filldraw[black] (-3,0) circle (2.5pt) node[above=4pt] {$M'$};
        \node[below=6pt] at (-3,0) {\small $-\frac{\pi}{8}$};

        \draw[latex-latex, thick, gray] (-3,-1.3) -- (3,-1.3) 
            node[midway, fill=white] {\small BZ Width $G = \frac{\pi}{4}$};
            
        \draw[dashed, gray] (-3,0) -- (-3,-1.4);
        \draw[dashed, gray] (3,0) -- (3,-1.4);

        \node[blue!80!black] at (1.5,0.4) {\scriptsize Non-HSP};
        \node[blue!80!black] at (-1.5,0.4) {\scriptsize Non-HSP};

\end{tikzpicture}
    \caption{Reduced $1D$ FBZ profile calculated for the $n=8$ sublattice configuration. High symmetry anchors lock at $k=0$ and the boundaries $k=\pm\pi/8$, while general interior wavevectors are unconstrained non-HSPs.}
    \label{fig:n8_brillouin_zone}
\end{figure}
\subsection{Case III: $\theta=\frac{\pi}{4}$} 
The unit cell now contains eight sublattices which results in more folded BZ and the size of the FBZ boundary further reduces $k\in [-\pi/8,~\pi/8]$ with the reduced reciprocal vector becoming $G=\pi/4$. We now have the reduced FBZ center is at $k=0~(\Gamma)$ and boundary is at $k=\pi/8~(M)$ and $-\pi/8~(M^{\prime})$. Therefore, the two HSP is visible here at $k_{0}=0$ and $k_{0}=\pm\pi/8$ which is shown schematically in Fig. \ref{fig:n8_brillouin_zone}.

\begin{figure}
   \vskip -.4 in
   \begin{picture}(100,100)
     \put(-70,0){
  \includegraphics[width=.45\linewidth, height=1.25 in]{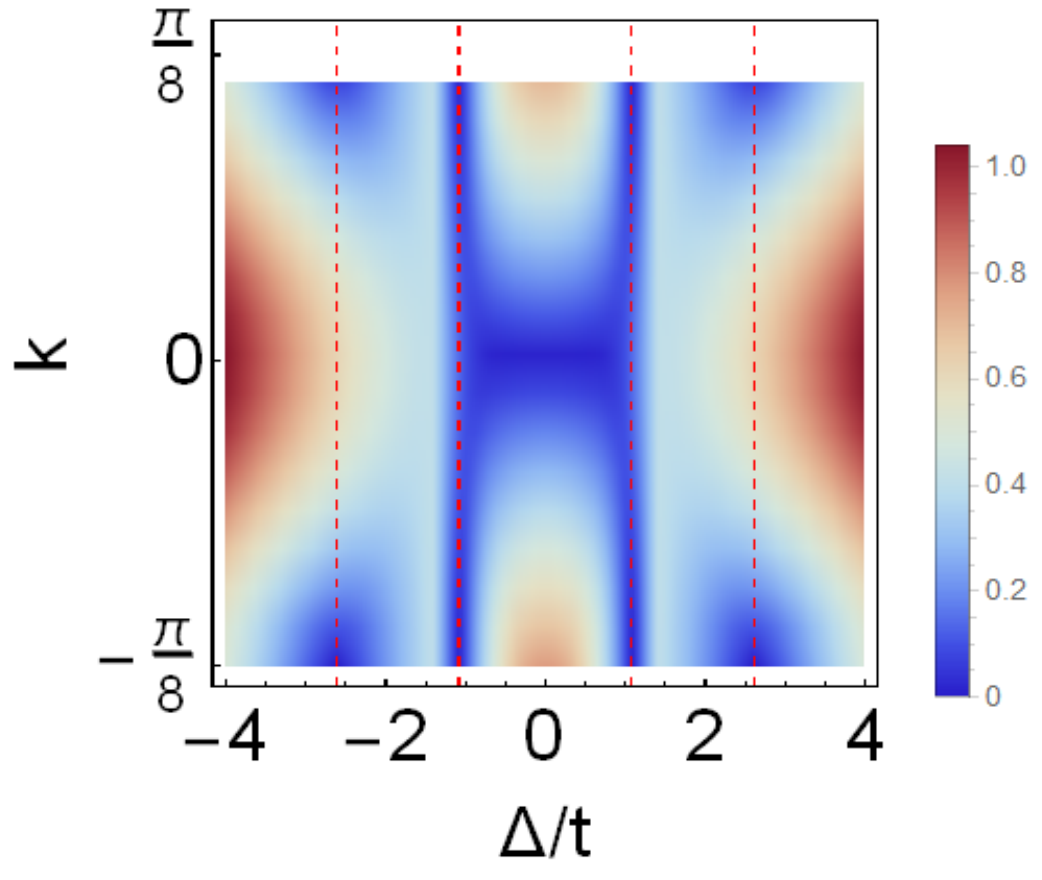}
  \includegraphics[width=.45\linewidth, height=1.25 in]{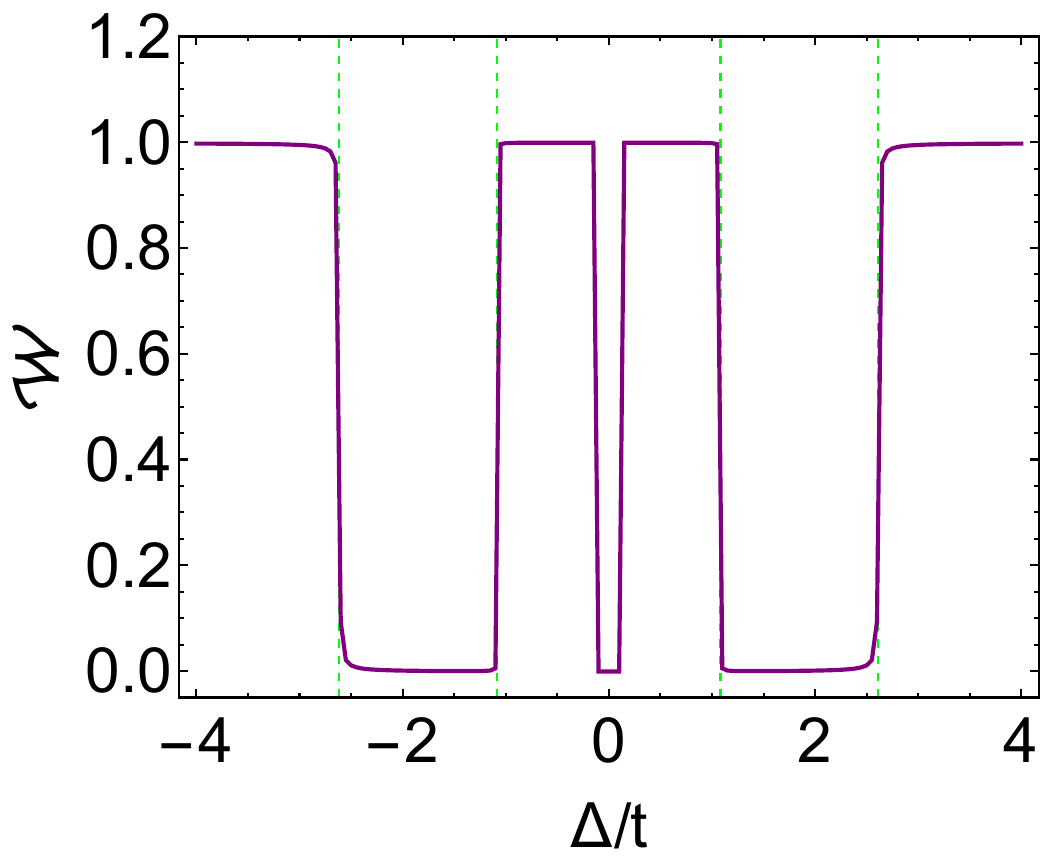}}
     \put(-10,62){(a)}
     \put(125,58){(b)}
    \end{picture}\\
     \vskip -.00004 in
   \begin{picture}(100,100)
     \put(-10,0){
       \includegraphics[width=.48\linewidth,height=1.25 in]{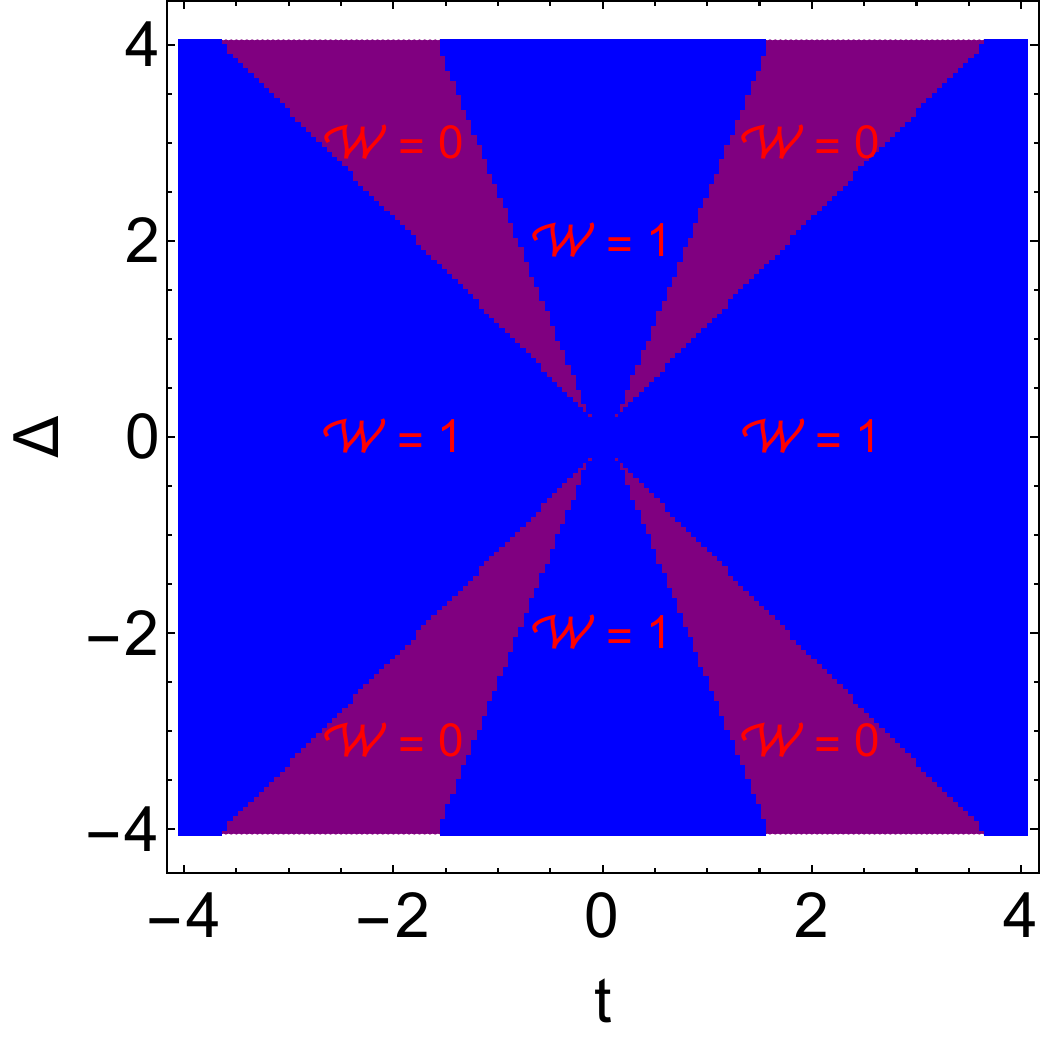}}
   \put(-10,58){(c)}
   \end{picture}
  \vskip -0.1 in
\caption{(a) Presents the band gap in the $(\Delta/t-k)$-plane for the SSH chain with eight sublattices in a unit cell. Here, $|\Delta/t|=\sqrt{2(2\pm\sqrt{2})}$ at HSP $k_0=\pm\pi/8$ and $\Delta=0$ at HSP $k_0=0$ are the gap closing line. The gap closing is indicated by the deep blue color in color bar. (b) The variation of winding number with $\Delta/t$. (c) The phase diagram in $(\Delta-t)$-plane where the topological region is marked by blue color and the trivial region is marked by purple color.} 
\label{figpiby4}
\end{figure}
\subsubsection{Hamiltonian and phase diagram}
In momentum space, this case gives $8\times 8$ Bloch Hamiltonian $H_{k}$ (for details see \cite{mandal}). The energy dispersion for this case can be estimated through
\begin{equation}\label{piby4}
E(k)_{\pm}=\pm\sqrt{t^4(t^2-\Delta^2)^2+A^2-2t^2(t^2-\Delta^2)A\cos8k}
\end{equation}
where, $A=t^4-t^2\Delta^2+\frac{\Delta^4}{4}$. Similar to $\theta=\pi/2$ case, here the above Eq. (\ref{piby4}) under PBC not only produces the gap closing transitions at $|\Delta/t|=\sqrt{2(2\pm\sqrt{2})}$ for HSP $k_0=\pm\pi/8$ but also gap closes at $\Delta=0$ for HSP $k_0=0$ which is illustrated in Fig. \ref{figpiby4}(a). The lowest positive energy band gap $E_{gap}\sim\Big|\sqrt{2(2\pm\sqrt{2})}t\pm\Delta\Big|$ can be discernible at the FBZ bounday  $k\in [-\pi/8,~\pi/8]$ for $|\Delta/t|\ne\sqrt{2(2\pm\sqrt{2})}$ while the energy band gap scale as $E_{gap}\sim|\Delta|^4$ at $k_0=0$ when $\Delta\ne0$. The lowest positive energy band exhibits Dirac-type linear band touching both under PBC and OBC at $|\Delta/t|=\sqrt{2(2\pm\sqrt{2})}$. Interestingly, it shows higher-order gap closing at $\Delta=0$ for the system with PBC (OBC), as here the band gap has the quartic ($|\Delta|^4$) dependence of parameter $\Delta$. Like $\theta=\pi/2$, here the winding number remains unchanged within a small region around $\Delta=0$, leading to a higher-order trivial topological transition. The winding number changes as one crosses $|\Delta/t|=\sqrt{2(2\pm\sqrt{2})}$ point, which gives rise to the Dirac-type gap-closing non-trivial topological transition. Therefore, the TPT may occur only at HSP $k_0=\pm\pi/8$ and here, like $\theta=\pi/2$, also HSP $k_{0}=0$ shows a higher-order trivial-to-trivial phase transition. Here, two ZES and six nonzero energy in-gap modes are found for the chain considering OBC\cite{mandal}. This case also falls in the $BDI$ class universality as the Hamiltonian respects chiral symmetry, time-reversal symmetry and particle-hole symmetry. An unitary transformation, however, makes the Hamiltonian $H_{k}$ into block off-diagonal form $H_{k}\rightarrow UH(k)U^{-1}=\begin{pmatrix}
0 & V \\
V^{\dagger} & 0 
\end{pmatrix}$ with 
\begin{equation}\label{32f}
V=
\begin{pmatrix}
 (t+\Delta) & 0 & 0 & (t+\frac{\Delta}{\sqrt{2}})e^{-8ik} \\
 (t+\frac{\Delta}{\sqrt{2}}) & t & 0 & 0 \\
 0 & (t-\frac{\Delta}{\sqrt{2}}) & (t-\Delta) & 0 \\
 0 & 0 & (t-\frac{\Delta}{\sqrt{2}}) & t
\end{pmatrix}.
\end{equation}

In order to distinguish the trivial and non-trivial topological phases, one needs to estimate the winding number given in Eq. (\ref{9}). One attains ${\mathcal W}=1$ for $0<|\Delta/t|<\sqrt{2(2-\sqrt{2})}$ or $|\Delta/t|>\sqrt{2(2+\sqrt{2})}$, signaling NTP and ${\mathcal W}=0$ when $\sqrt{2(2-\sqrt{2})}<|\Delta/t|<\sqrt{2(2+\sqrt{2})}$ or $0\le|\Delta/t|<<\sqrt{2(2-\sqrt{2})}$ indicating TTP. Consequently, the four TPT is noticeable here at the critical point $|\Delta/t|=\sqrt{2(2\pm\sqrt{2})}$. The winding for different phases is displayed in Fig. \ref{figpiby4}(b), which includes a trivial-to-trivial phase transition within the range  $0\le|\Delta/t|<<\sqrt{2(2-\sqrt{2})}$. Fig. \ref{figpiby4}(c) exhibits the corresponding phase diagram.

This is a multiband model, like $\theta=\pi/2$. Thus, the curvature function estimated from Eq. (\ref{9}) to be
\begin{widetext}
\begin{equation}\label{curvaturepiby4}
F(k,{\bf M})=\frac{d\phi_{k}}{dk}=\frac{d}{dk} \arctan\Big(\frac{Im[{\bf Det[V(k)]]}}{Re[{\bf Det[V(k)]}]}\Big)=\frac{1}{\frac{1}{4}-\frac{\Delta^4(\Delta^4-8t^2\Delta^2+8t^4)}{8(\Delta^2-2t^2)^2\Big((\Delta^2-2t^2)^2+4(\Delta-t)t^2+t^2(\Delta+t)\cos8k\Big)}}
\end{equation} 
\end{widetext}

\begin{figure}
   \vskip -.4 in
   \begin{picture}(100,100)
     \put(-70,0){
  \includegraphics[width=.45\linewidth, height=1.25 in]{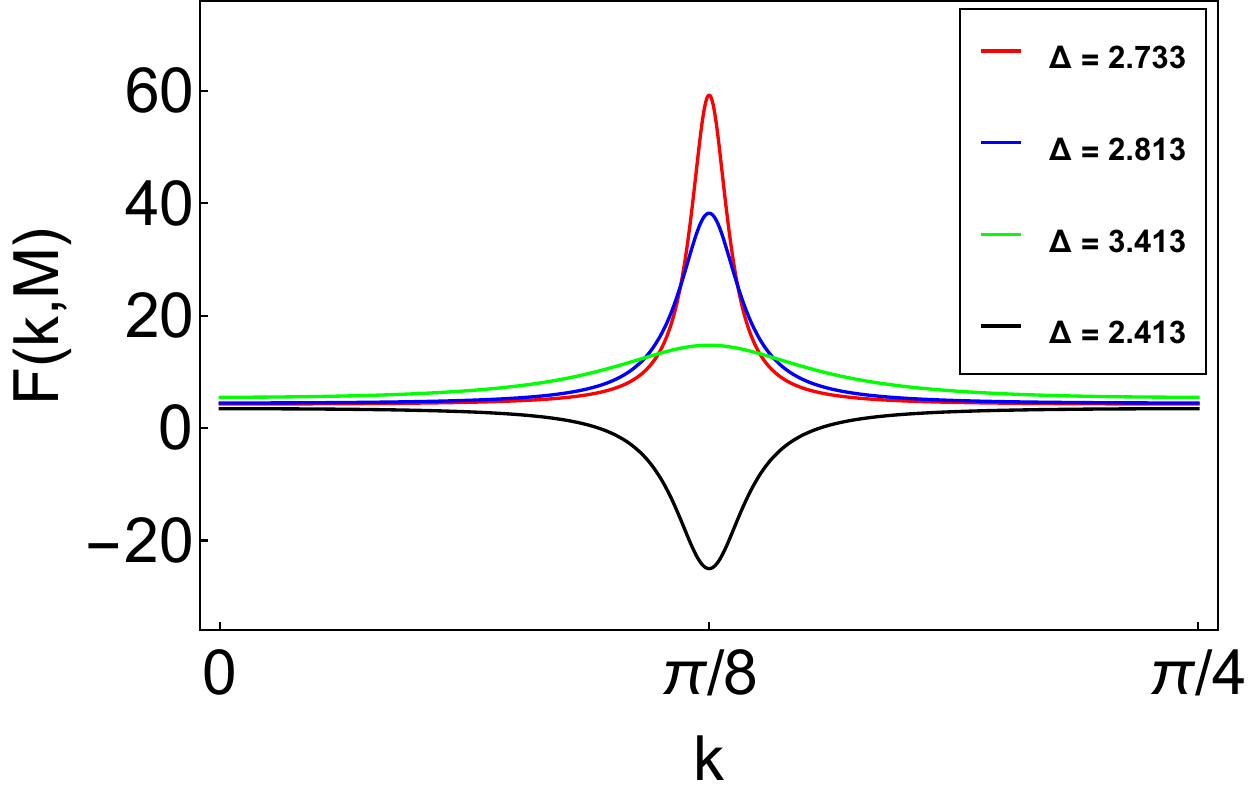}
  \includegraphics[width=.45\linewidth, height=1.25 in]{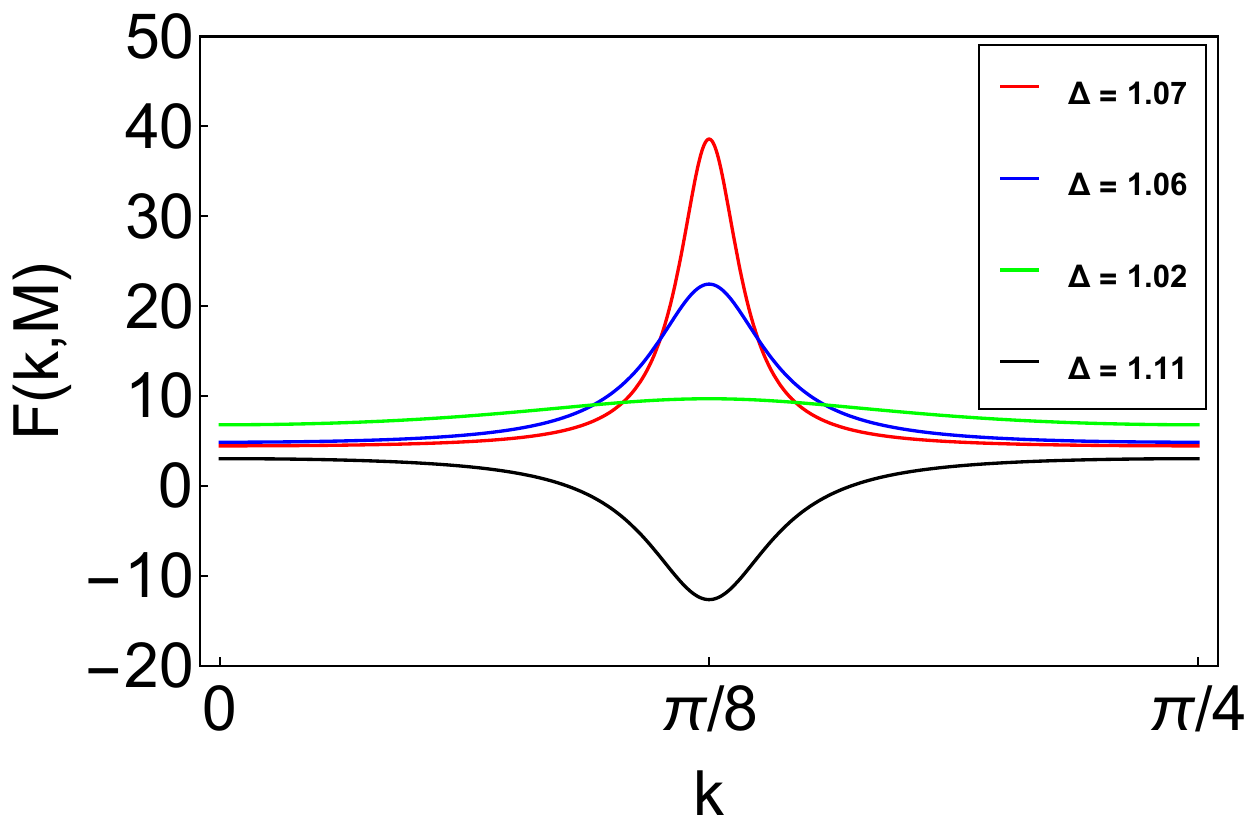}}
     \put(-30,62){(a)}
     \put(115,58){(b)}
    \end{picture}\\
     \vskip -.00004 in
   \begin{picture}(100,100)
     \put(-10,0){
       \includegraphics[width=.48\linewidth,height=1.25 in]{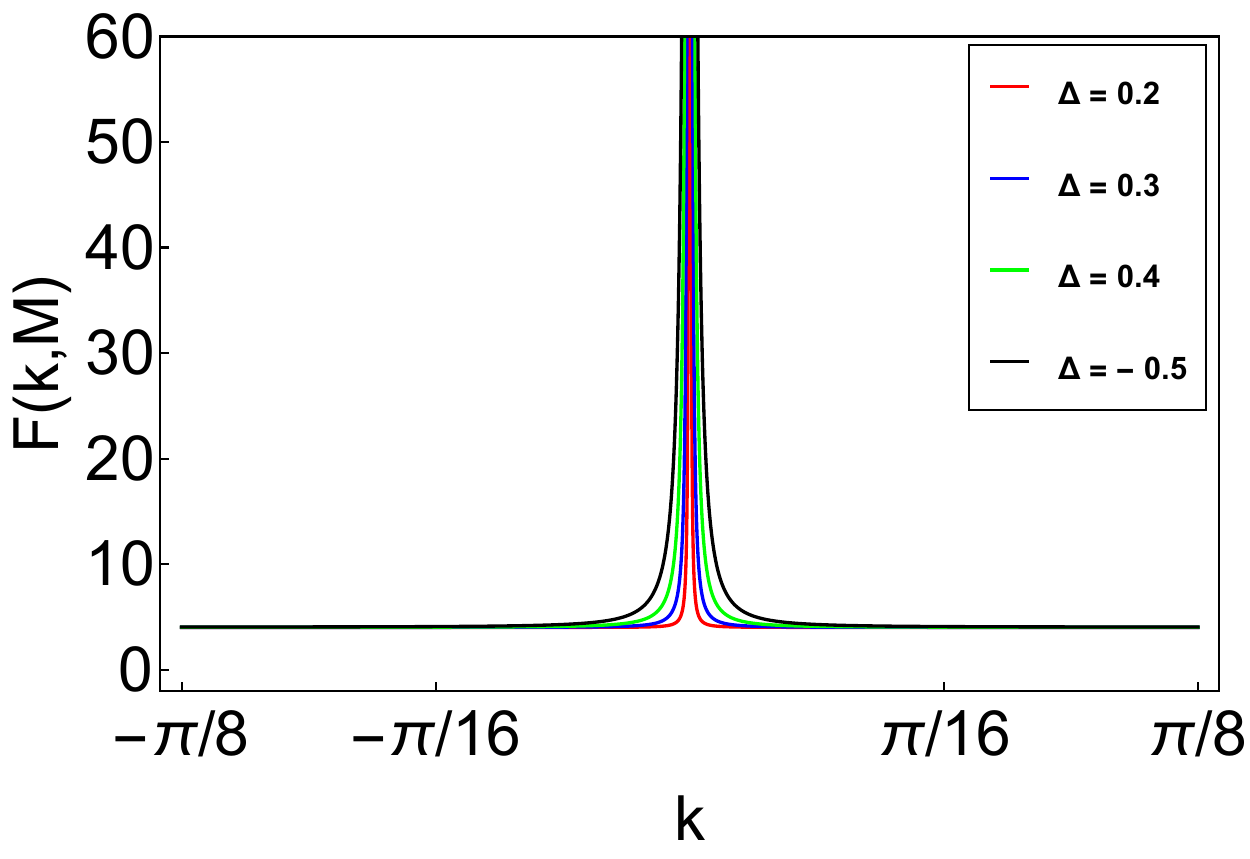}}
   \put(30,58){(c)}
   \end{picture}
  \vskip -0.1 in
\caption{(a) and (b) The curvature function $F(k,{\bf M})$ around HSP $k_{0}=\pi/8$. (c) The same near HSP $k_{0}=0$ which exhibits faster divergence.} 
\label{figfk}
\end{figure}
Just as before, here is how the curvature function diverges near the four non-trivial topological transition points and near the higher order trivial transition point is depicted in Fig. \ref{figfk}. Like the previous trend, here the divergence of $F(k,{\bf M})$ increases as one moves towards the critical point and away from it give rise to divergence reduction. Looking carefully at the Figs. \ref{figfk}(a) and (b) reveals that $\Delta(=2.733)\rightarrow\Delta^{\prime}(=2.813)$ and $\Delta(=1.07)\rightarrow\Delta^{\prime}(=1.06)$ are the CRG flow near the critical points $|\Delta/t|=\sqrt{2(2+\sqrt{2})}$ and $|\Delta/t|=\sqrt{2(2-\sqrt{2})}$, respectively,  through which the divergence reduction is visible. Here, one can also notice faster divergence of $F(k,{\bf M})$ near the higher-order gap closing point as compared to the non-trivial topological transition points (see Fig. \ref{figfk}(c)). Importantly, this divergence is steeper than that noticed near the trivial metallic point for $\theta=\pi/2$ due to the fact that the energy gap scales here as $|\Delta|^{4}$.

\subsubsection{Flow equations, fixed points and critical points}
In order to apply the CRG procedure, we insert Eq. (\ref{curvaturepiby4}) into Eq. (\ref{curvature5}) and calculate the RG flow equations around $k_{0}=\pi/8$ for the parameters ${\bf M}=(t,\Delta)$ as
\begin{equation}\label{rg6}
\frac{d\Delta}{dl}=4\Big(\Delta+\frac{(3\Delta^5-4\Delta^3t^2)}{(\Delta^4-8\Delta^2t^2+8t^4)}\Big),
\end{equation}

\begin{equation}\label{rg7}
\frac{dt}{dl}=-\frac{16t(\Delta^4-3\Delta^2t^2+2t^4)}{(\Delta^4-8\Delta^2t^2+8t^4)}
\end{equation}

\begin{figure}
   \vskip -.4 in
   \begin{picture}(100,100)
     \put(-70,0){
  \includegraphics[width=.45\linewidth, height=1.25 in]{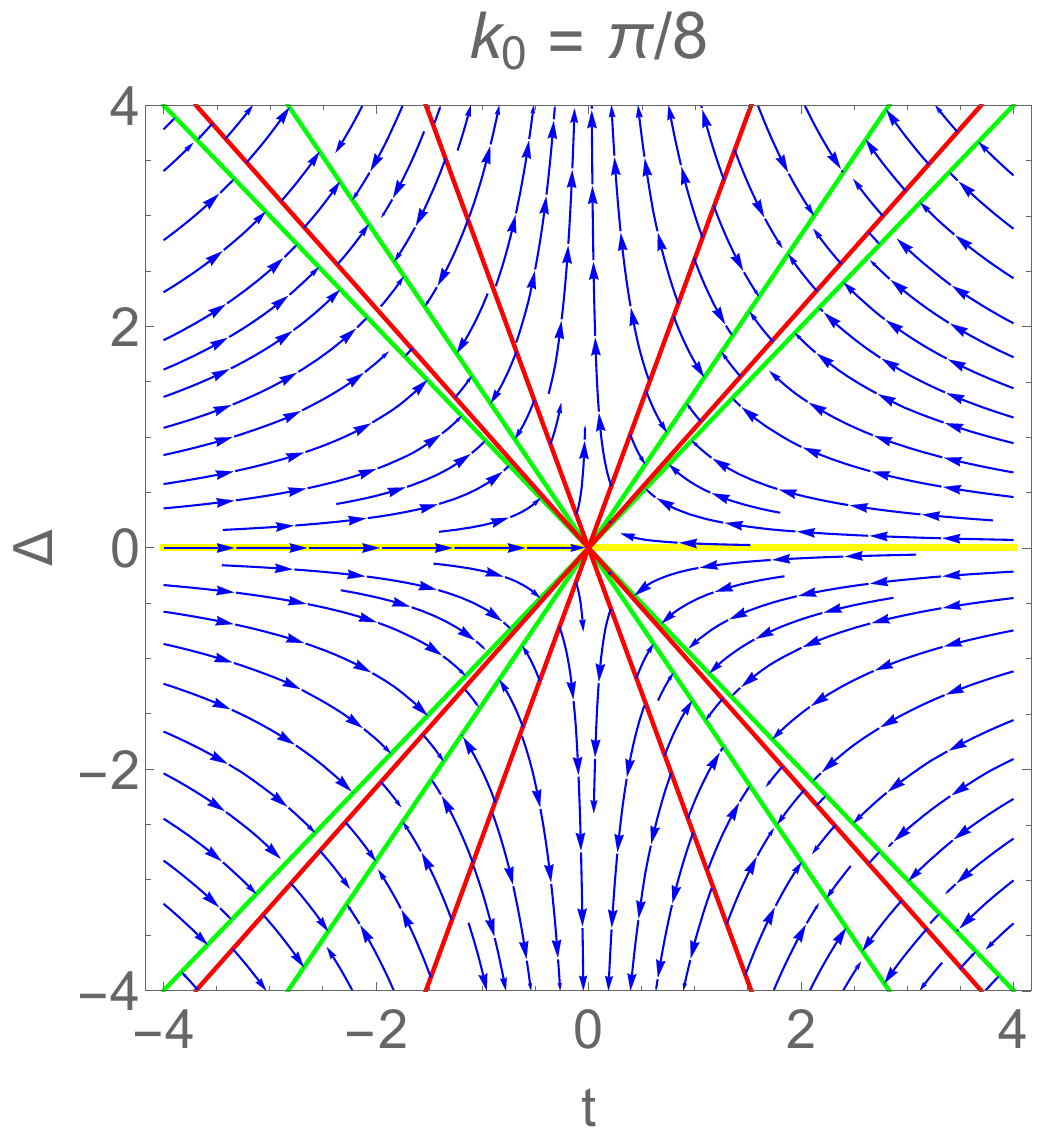}
  \includegraphics[width=.45\linewidth, height=1.25 in]{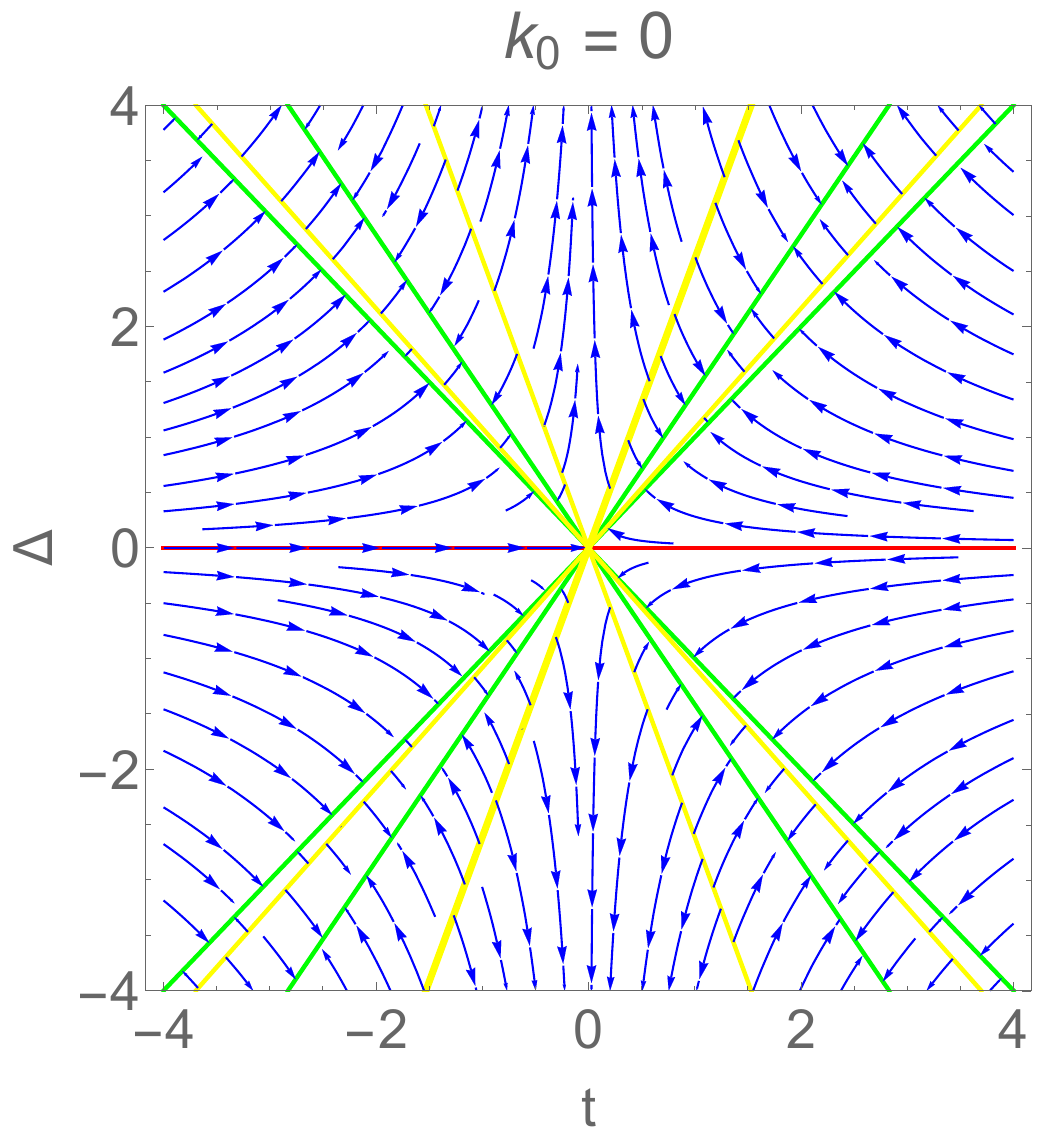}}
     \put(10,62){(a)}
     \put(125,58){(b)}
    \end{picture} 
  \vskip -0.1 in
\caption{ The CRG flow diagram near HSP $k_{0}=\pi/8$ (a) and $k_{0}=0$ (b). The red line in (a) indicates the topological critical points, while the same in (b) represents the higher-order non-topological transition point. The green lines and yellow lines both in (a) and (b) are for fixed points.} 
\label{figflow}
\end{figure}

Moreover, the flow equations near $k_{0}=0$ is estimated as
\begin{equation}\label{rg8}
\frac{d\Delta}{dl}=\frac{16(\Delta^2-t^2)(\Delta^2-2t^2)(\Delta^4-8\Delta^2t^2+8t^4)}{\Delta^7},
\end{equation}

\begin{equation}\label{rg9}
\frac{d\Delta}{dl}=-\frac{16t(\Delta^2-t^2)(\Delta^2-2t^2)(\Delta^4-8\Delta^2t^2+8t^4)}{\Delta^8}
\end{equation}
Now, like the previous cases, the flows of ($t,\Delta$) as obtained from CRG Eqs. (\ref{rg6})-(\ref{rg9}) are presented in Fig. \ref{figflow}. Again, it is noticeable that the flow diagram correctly produces the phase diagram Fig. \ref{figpiby4}(c). From Eqs. (\ref{rg6})-(\ref{rg9}), we can correctly estimate the critical points at $|\Delta/t|=\sqrt{2(2\pm\sqrt{2})}$ and higher-order trivial point at $\Delta=0$. Moreover, the fixed points are obtained at $\Delta=0, \pm t, \pm\sqrt{2}t$ for $k_{0}=\pi/8$ and it is at $\Delta=\pm t, \pm\sqrt{2}t,\pm\sqrt{2(2\pm\sqrt{2})}t$ when $k_{0}=0$. The fixed points at $\Delta=\pm t$ are topologically non-trivial, whereas those at $\Delta=\pm \sqrt{2}t$ are topologically trivial. Further, we can observe here that the flow rate diverges along the critical line and higher-order trivial transition line, which are presented by red lines in Figs. \ref{figflow}(a) and (b). In those plots, the fixed points for both cases are displayed by green and yellow lines. The flow rate converges (diverges) for green (yellow) lines, indicating the stable (unstable) fixed points. 

From Fig. \ref{figflow}(a), it is noticed that the RG flow diagram clearly reproduces the $k_{0}=\pi/8$ critical line, and the unstable fixed line also reproduces the $k_{0}=0$ second-order trivial transition line almost completely. In the same manner, Fig. \ref{figflow}(b) shows the reproduction of $k_{0}=0$ second-order trivial transition line along with the regenerating of the $k_{0}=\pi/8$ critical line.

Now, it is worth emphasizing that the RG flows near a Dirac-like HSP are also able to reproduce the phase boundaries around non-trivial higher-order HSP. Importantly, the RG flow equations near a point $k_{0}$ are sufficient to predict the phase boundaries in the entire BZ\cite{rao}.

\subsubsection{Correlation length and critical exponents}
Here, we calculate the curvature function $F(k,\bf{M})$ with the help of Eq. (\ref{curvaturepiby4}) and takes the following form:
\begin{equation}\label{co4a}
F(k_{0},{\bf M})=\begin{cases}
{\frac{8(\Delta^2-2t^2)^2}{\Delta^4-8\Delta^2t^2+2t^4}}, & \text{at $k_{0}=\pi/8$}\\
~~~~~~\frac{8(\Delta^2-2t^2)}{\Delta^4}, & \text{at $k_{0}=0$}
\end{cases}
\end{equation}

The divergence of $F(k_{0},{\bf M})$ near the criticality $|\Delta_{c}|=\sqrt{2(2\pm\sqrt{2})}t$ and gap closing points $\Delta_{c_{2}^{\prime}}=0$ scale as
\begin{equation}\label{co5a}
F(k_{0},{\bf M})\propto\begin{cases}
{~~|\Delta^2-\Delta_{c}^2|^{-1}}, & \text{near $\Delta_{c}$}\\
~~|\Delta-\Delta_{c_{2}^{\prime}}|^{-4}, & \text{near $\Delta_{c_{2}^{\prime}}$}
\end{cases}
\end{equation}
the above Eq. (\ref{co5a}) giving $\gamma=1$ ($\gamma=4$) for $\Delta_{c}$ ($\Delta_{c_{2}^{\prime}}$).

The correlation length $\xi_{k_{0}}$, close to the criticality, estimated as

\begin{equation}\label{co6a}
\xi_{k_{0}}({\bf M})=\begin{cases}
{~~\sqrt{32-\frac{96A}{B}+\frac{64A^2}{B^2}}}~~~\stackrel{\Delta \to \Delta_{c}}{\propto} ~~|\Delta^2-\Delta_{c}^2|^{-1}\\
~\sqrt{32+\frac{96B}{C-B}+\frac{64B^2}{(C^2-B^2)^2}}~\stackrel{\Delta \to \Delta_{c_{2}^{\prime}}}{\propto}|\Delta-\Delta_{c_{2}^{\prime}}|^{-4}
\end{cases}
\end{equation}
with $A=2t^4-2\Delta^2t^2+\frac{\Delta^4}{4}$, $B=t^4-\Delta^2t^2+\frac{\Delta^4}{4}$, and $C=t^2(t^2-\Delta^2)$. Eq. (\ref{co6a}) provides $\nu=1$ ($\nu=4$) for $\Delta_{c}$ ($\Delta_{c_{2}^{\prime}}$).

Following Eqs. (\ref{co5a}) and (\ref{co6a}), we see the linear behavior of $F(k_{0},{\bf M})$ and $\xi_{k_{0}}({\bf M})$ when approaching the critical phase boundary $\Delta_{c}$ from either direction. However, a quartic type behavior is observed as approaching $\Delta_{c_{2}^{\prime}}$.
\begin{figure}
   \vskip -.1 in
   \begin{picture}(100,100)
     \put(-70,0){
  \includegraphics[width=.45\linewidth, height=1.25 in]{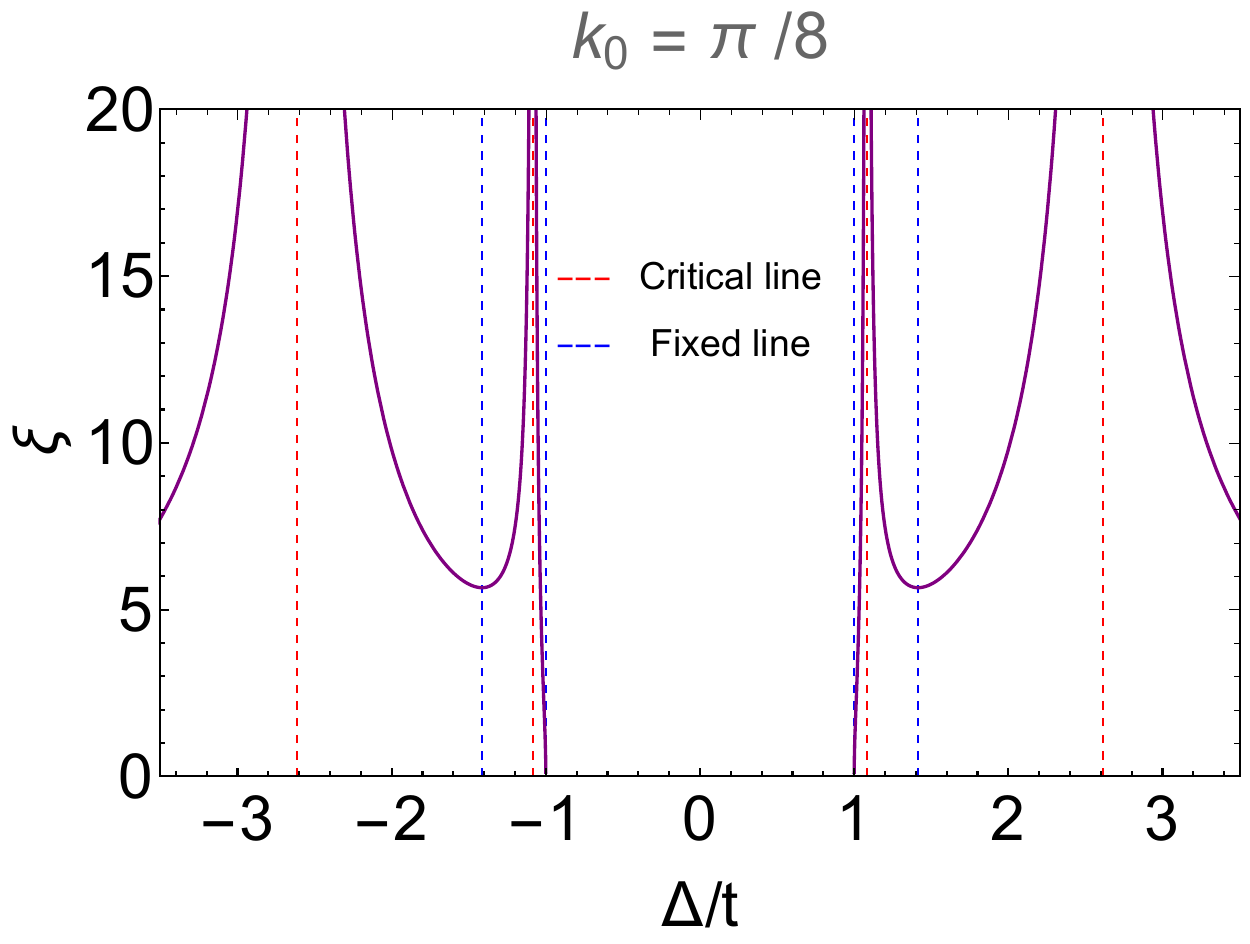}
  \includegraphics[width=.45\linewidth, height=1.25 in]{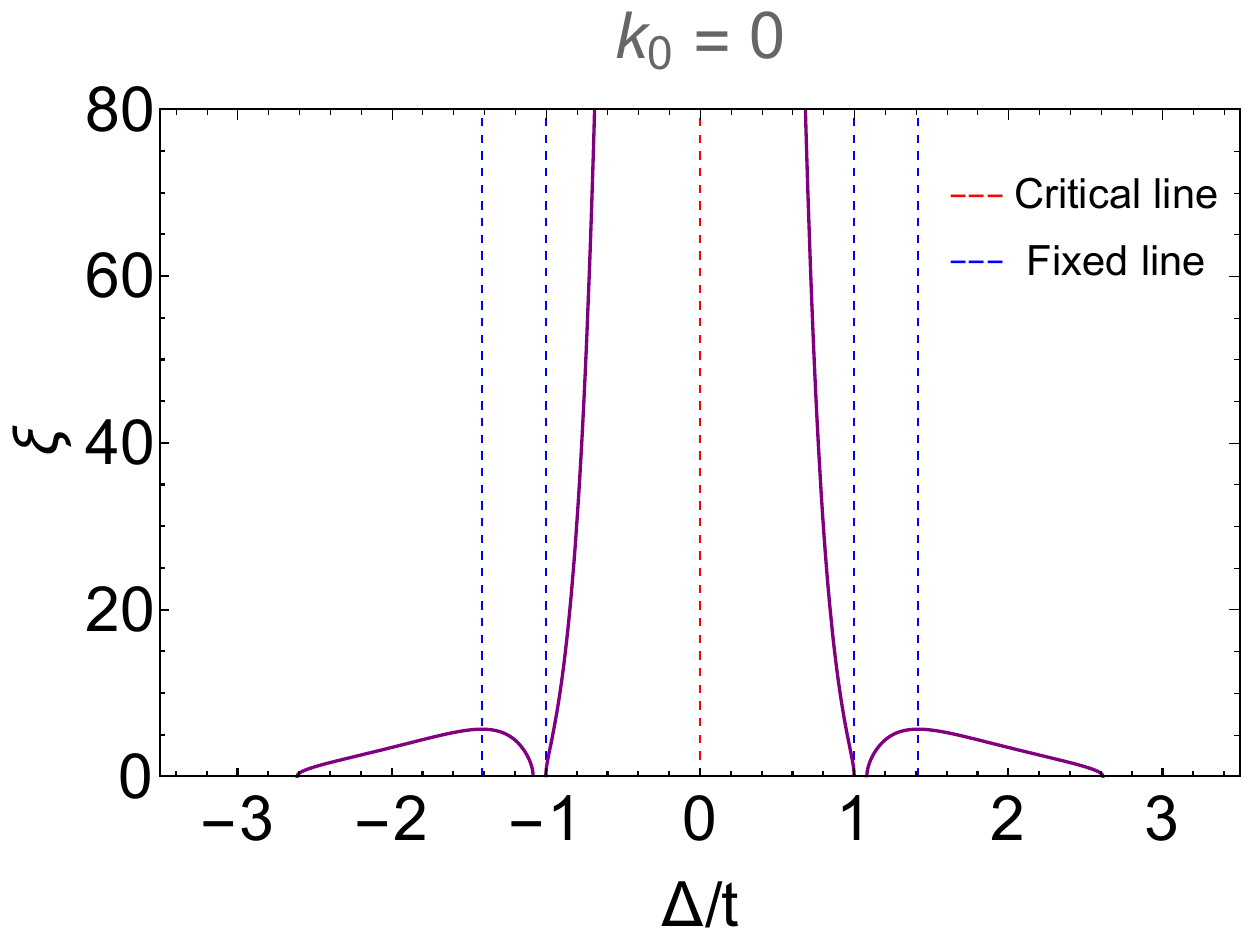}}
     \put(-15,62){(a)}
     \put(70,62){(b)}
    \end{picture} 
  \vskip -0.1 in
\caption{ The correlation length against $\Delta/t$ with $t=1$ for (a) $k_{0}=\pi/8$ and (b) $k_{0}=0$.} 
\label{figcorre}
\end{figure}

We now plot Fig. \ref{figcorre} to demonstrate the behavior of correlation length for $k_{0}=\pi/8$ and $k_{0}=0$. The plot depicts the divergence of $\xi_{k_{0}}$ at $\Delta=\pm t,\pm \sqrt{2}t$ (Fig. \ref{figcorre}(a)) and the divergence becomes relatively faster as $\Delta\rightarrow0$ (Fig. \ref{figcorre}(b)). In both plots,  we notice the vanishing of $\xi_{k_{0}}$ at non-trivial fixed points $\Delta=\pm t$ while there is a finite value of $\xi_{k_{0}}=4\sqrt{2}$ at the trivial fixed points $\Delta=\pm\sqrt{2}t$.

\subsubsection{Edge-state localization at the critical and fixed points}
As in the previous two cases, the edge modes here exhibit complete delocalization at the critical and higher-order gap-closing points, namely at $|\Delta/t|=\sqrt{2(2\pm\sqrt{2})}$ and $\Delta=0$. Starting from $|\Delta/t|=\sqrt{2(2+\sqrt{2})}$ and moving towards the trivial fixed points, the localization of the edge modes progressively increases. Upon crossing these trivial fixed points, the modes become completely delocalized at $|\Delta/t|=\sqrt{2(2-\sqrt{2})}$. Conversely, as one moves from $|\Delta/t|=\sqrt{2(2-\sqrt{2})}$ towards the non-trivial fixed points, the edge modes again exhibit increasing localization, reaching their maximum localization at the non-trivial fixed points. Finally, upon moving away from these non-trivial fixed points towards $\Delta=0$, the localization gradually decreases.
\begin{table*}
\parbox{.82\linewidth}{
\centering
\begin{tabular}{  p{1.5cm}| p{3cm}| p{4cm}|p{2cm} |p{2cm} } 
 \hline
 {\bf$\theta$ values} & {\bf Band gap scaling} & {\bf Gap closing nature}& {\bf$\nu$} & {\bf$\gamma$} \\ [0.65ex] 
 \hline
 $\pi$ & $E_{gap}\sim |\Delta|^{1}$ & Linear or Dirac-type&1 & 1 \\ [1.2ex] 
 \hline
  $\pi/2$ & $E_{gap}\sim |\Delta|^{2}$ & Quadratic type&2 & 2 \\ [1.2ex] 
 \hline
  $\pi/4$ & $E_{gap}\sim |\Delta|^{4}$ & Quatic type&4 & 4 \\[1.2ex] 
 \hline
\end{tabular}
\caption{Comparison of critical exponents near $\Delta=0$ for different $\theta$ values.}
\label{table:1}}
\end{table*}

\begin{figure}
   \vskip -.4 in
   \begin{picture}(100,100)
     \put(-30,0){
  \includegraphics[width=.65\linewidth]{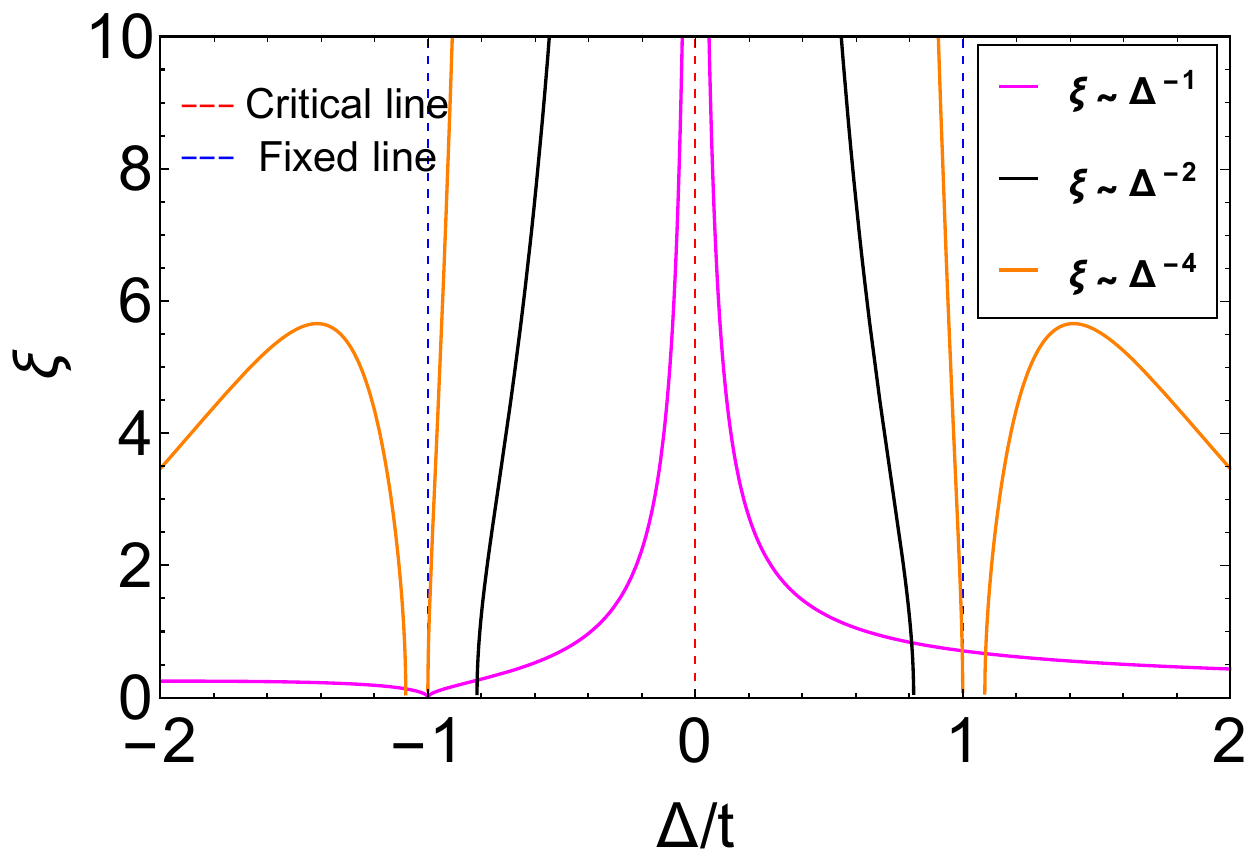}}
     \put(1,62){}
   \end{picture} 
\caption{The comparative plot for divergence of correlation length at $k_{0}=0$ for $\Delta=0$.} 
\label{figcompa}
\end{figure}

Lastly, we summarize the band gap scaling and critical exponents for different $\theta$ values around the point $\Delta=0$ in Table.\ref{table:1}. Therefore, we can conclude here that one can get higher critical exponents near this point, which results in faster divergence of the correlation length, as the periodicity of the system increases (see Fig. \ref{figcompa}).

\section{Summary}\label{sec4}

Based on the CRG analysis, in this paper we have investigated in detail the nature of transitions, localization of topological zero energy modes and the degree of divergence of curvature functions at the gapless points in a SSH model spectra in presence of periodic hopping modulations. The standard SSH chain shows a TPT at $\Delta=0$ where the curvture function diverges linearly with $\Delta$. However, for periodic hoppig modulation with $\theta=\pi/2$ or $\pi/4$, we observe the $\Delta=0$ point, which no more remains a point of TPT but still a gapless point, to exhibit a quadratic and quartic dependence of $\Delta$ respectively in the divergent curvature function there. On the contrary, the points of TPT in those respective cases continue to show linear Dirac-like divergences in $F(k_0,{\bf M})$ characterising slower divergences. Hence the topological and non-topological phase transitions that we encounter here fall under different universality classes with different sets of critical exponents. Interestingly, the edge localization of zero energy topological modes get completely disrupted giving fully delocalized modes due to diverging correlation (or, decay) length appearing at the point of TPT. On the other hand, these modes get fully localized at the notrivial fixed points (within the topological regime). We find that such fixed points appear at $\Delta=-t,~\pm t$ and $\pm t$ for $\theta~=~\pi,~\pi/2$ and $\pi/4$ respectively.

Thus using renormalization of curvature functions, this study uncovers important details of the quantum criticalities in the topologically relevant SSH model and its periodically hopping modulated variants. One can even explore the critical dynamics of related periodically driven Floquet systems in the similar fashion utilizing the discretized RG equations\cite{rg1}.

\section*{Acknowledgements}
 This work is financially supported by DST-SERB, Government of India via grant no. CRG/2022/002781.

\appendix
\section{Quadratic Gap-Closing for a Periodic Chain}\label{gap-quadratic}
\begin{figure}
   \begin{picture}(100,100)
     \put(-30,0){
  \includegraphics[width=.6\linewidth]{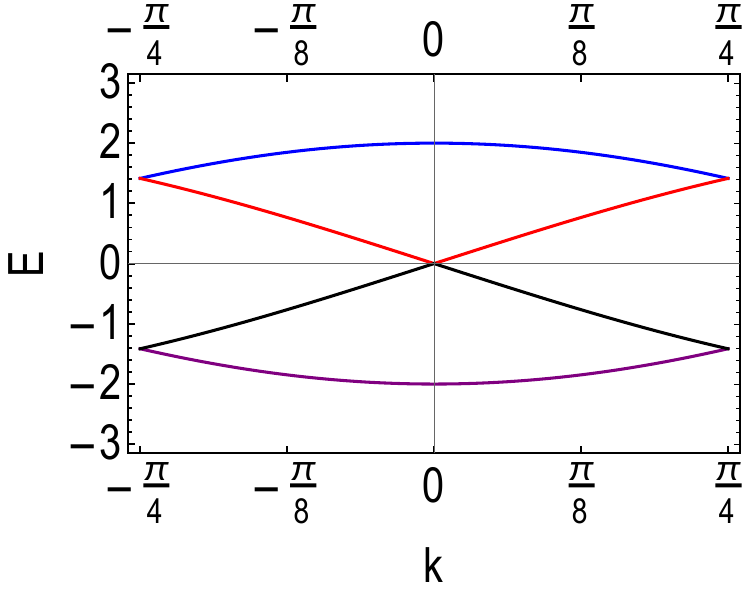}}
     \put(1,62){}
   \end{picture} 
\caption{The quadratic-type gap-closing for a four-site periodic SSH chain at $k_{0}=0$ for $\Delta=0$.} 
\label{figquadratic}
\end{figure}

For PBC, the SSH chain shows a Dirac-type (linear) band touching at $|\Delta/t|=\sqrt{2}$ \cite{mandal}, while it shows quadratic-type band touching at $\Delta=0$ as presented in Fig. \ref{figquadratic}.
\section{Graphical nature of edge-mode localization at the critical and fixed points}\label{critical-fixed}
\begin{figure}
   \begin{picture}(100,100)
     \put(-70,0){
  \includegraphics[width=.45\linewidth, height=1.25 in]{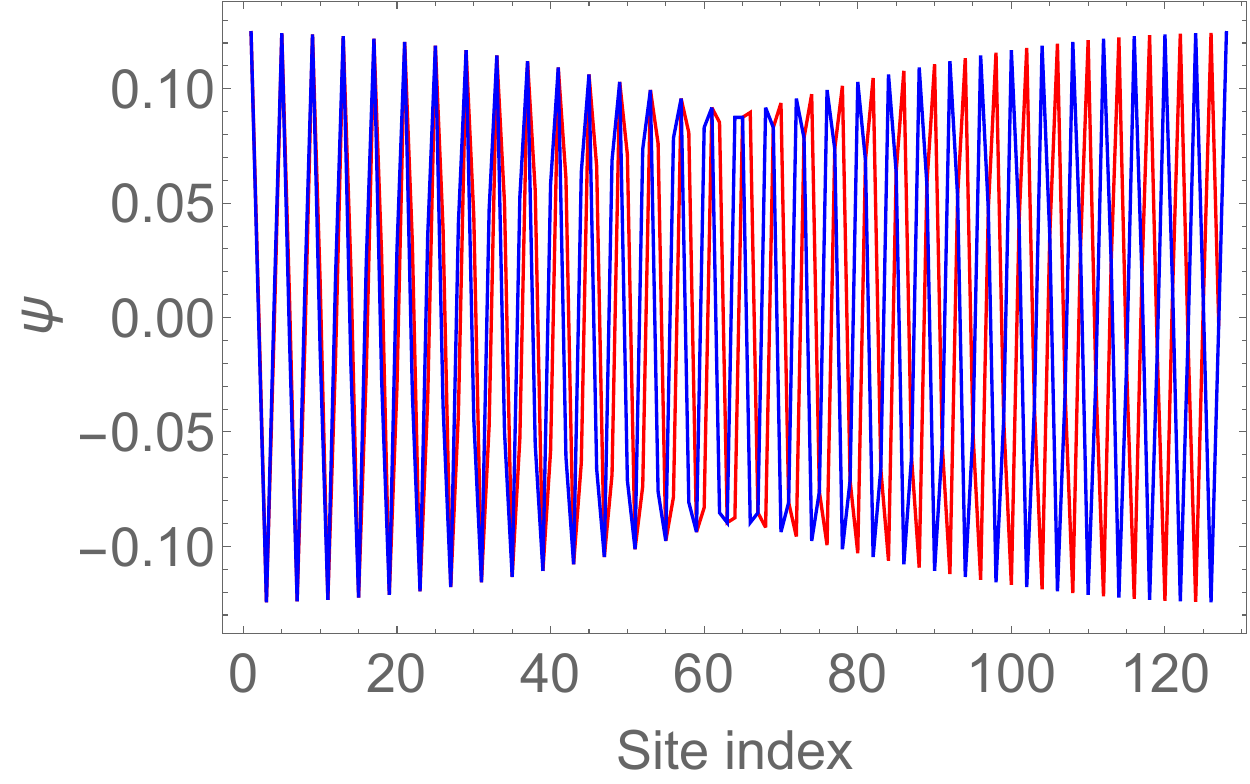}
  \includegraphics[width=.45\linewidth, height=1.25 in]{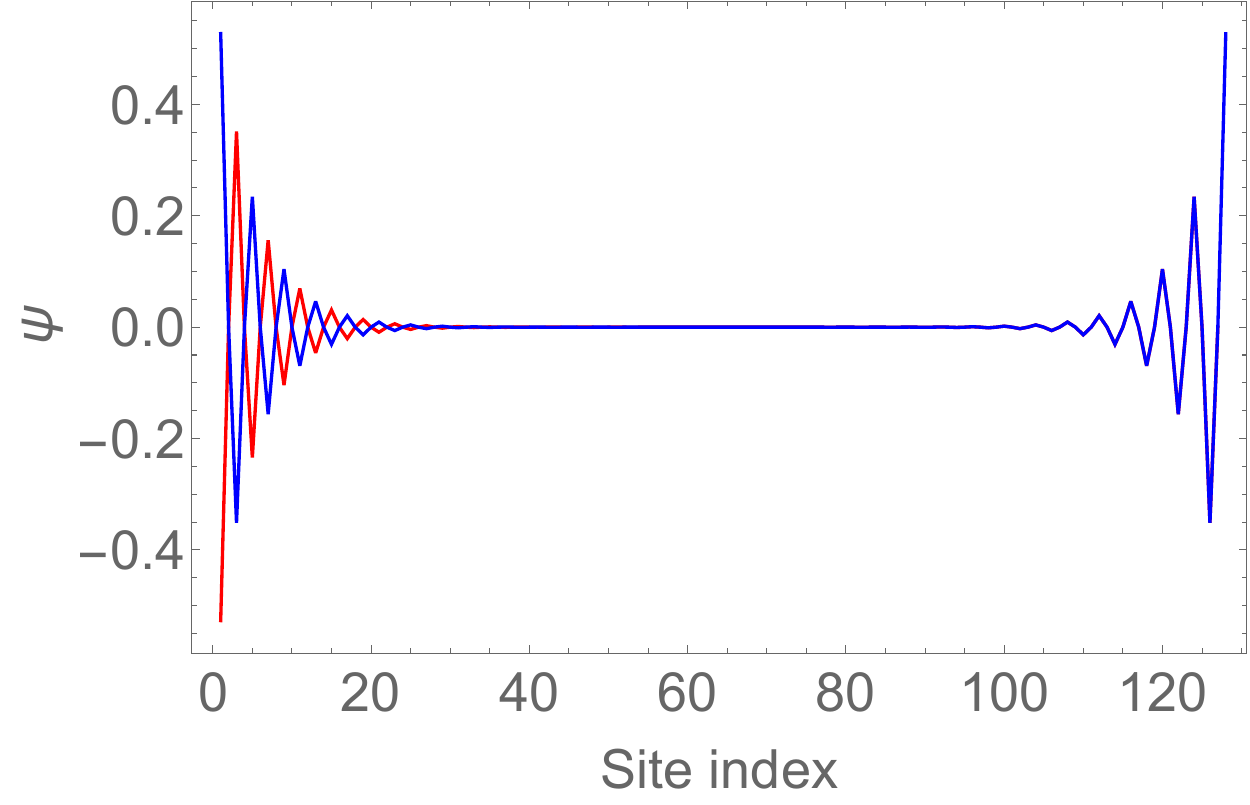}}
     \put(-10,62){(a)}
     \put(125,58){(b)}
    \end{picture}\\
     \vskip -.00004 in
   \begin{picture}(100,100)
     \put(-70,0){
    \includegraphics[width=.45\linewidth,height=1.25 in]{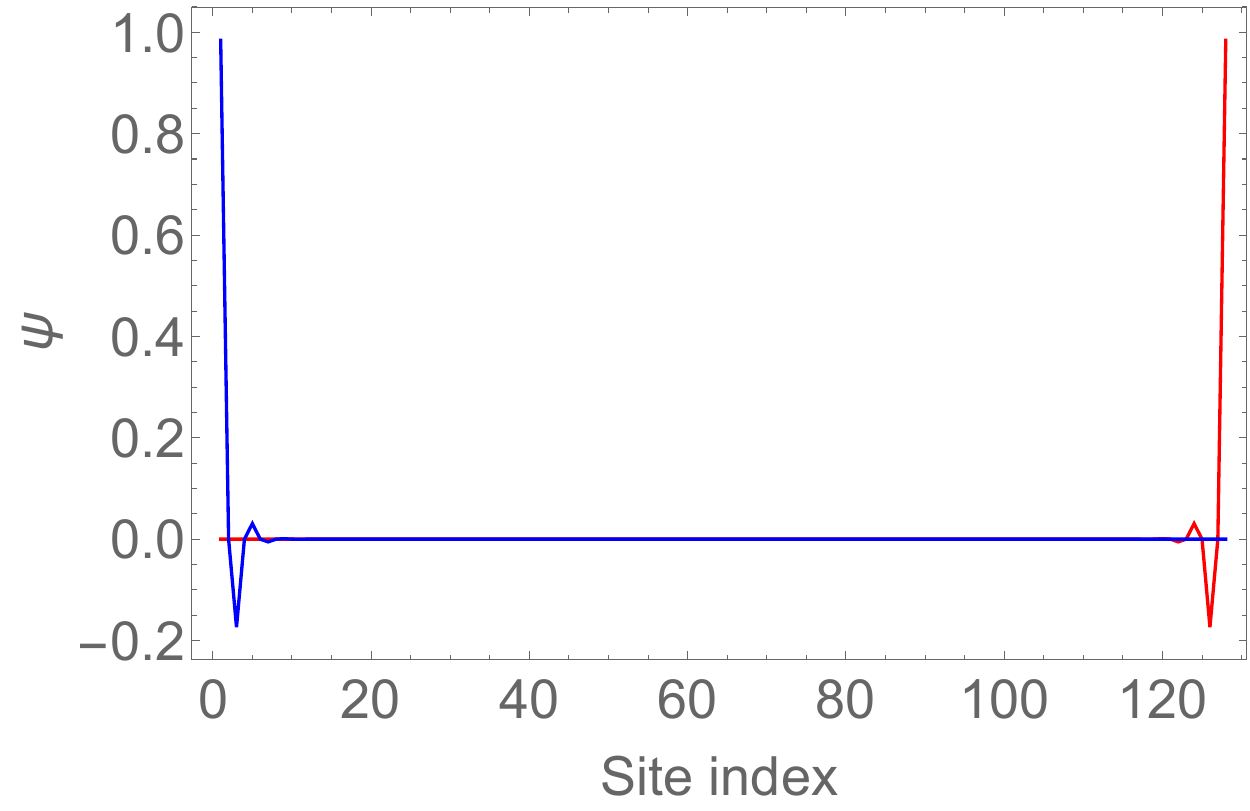}
      \includegraphics[width=.45\linewidth,height=1.25 in]{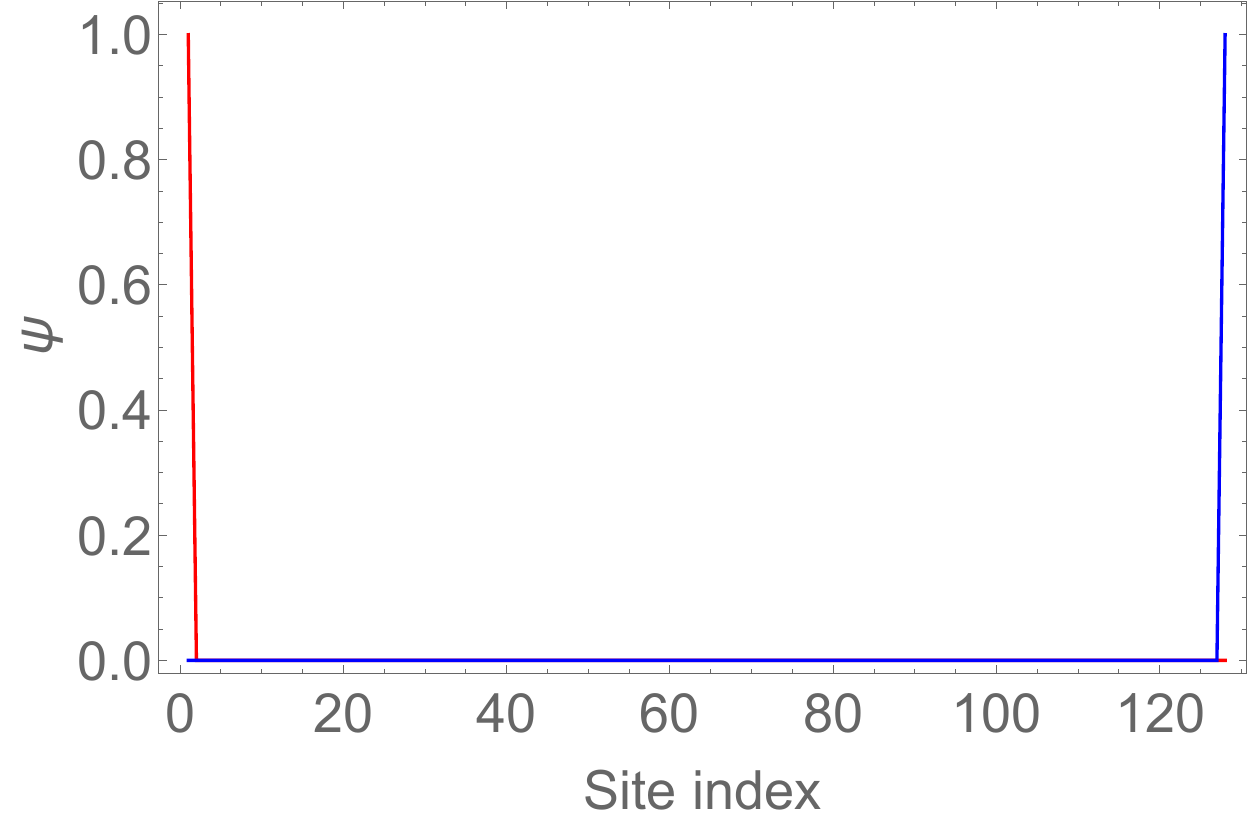}}
   \put(-10,62){(c)}
    \put(125,58){(d)}
   \end{picture}
  \vskip -0.1 in
\caption{ Behavior of edge modes for different choices of $\Delta$ for $\theta=\pi$. In (a) $\Delta=0$, (b) $\Delta/t=-0.2$, (c) $\Delta/t=-0.7$, and (d) $\Delta/t=-1$.} 
\label{figedge-1}
\end{figure}

Here, we intend to plot the behavior of the edge modes near the critical and fixed points for $\theta=\pi$ in Fig. \ref{figedge-1}. It is noticeable from Fig. \ref{figedge-1}(a) that the edge states exhibit complete delocalization at the critical point. As the system moves away from criticality into the topological regime, the edge states become progressively more localized, characterized by an increasingly rapid decay of their amplitudes into the bulk (notice Figs. \ref{figedge-1}(b) and (c)). The localization reaches its maximum at the nontrivial fixed point, $\Delta=-t$, where the edge modes are sharply confined to the boundary (see Fig. \ref{figedge-1}(d)).

\begin{figure}
   \vskip -0 in
   \begin{picture}(100,100)
     \put(-70,0){
  \includegraphics[width=.45\linewidth, height=1.25 in]{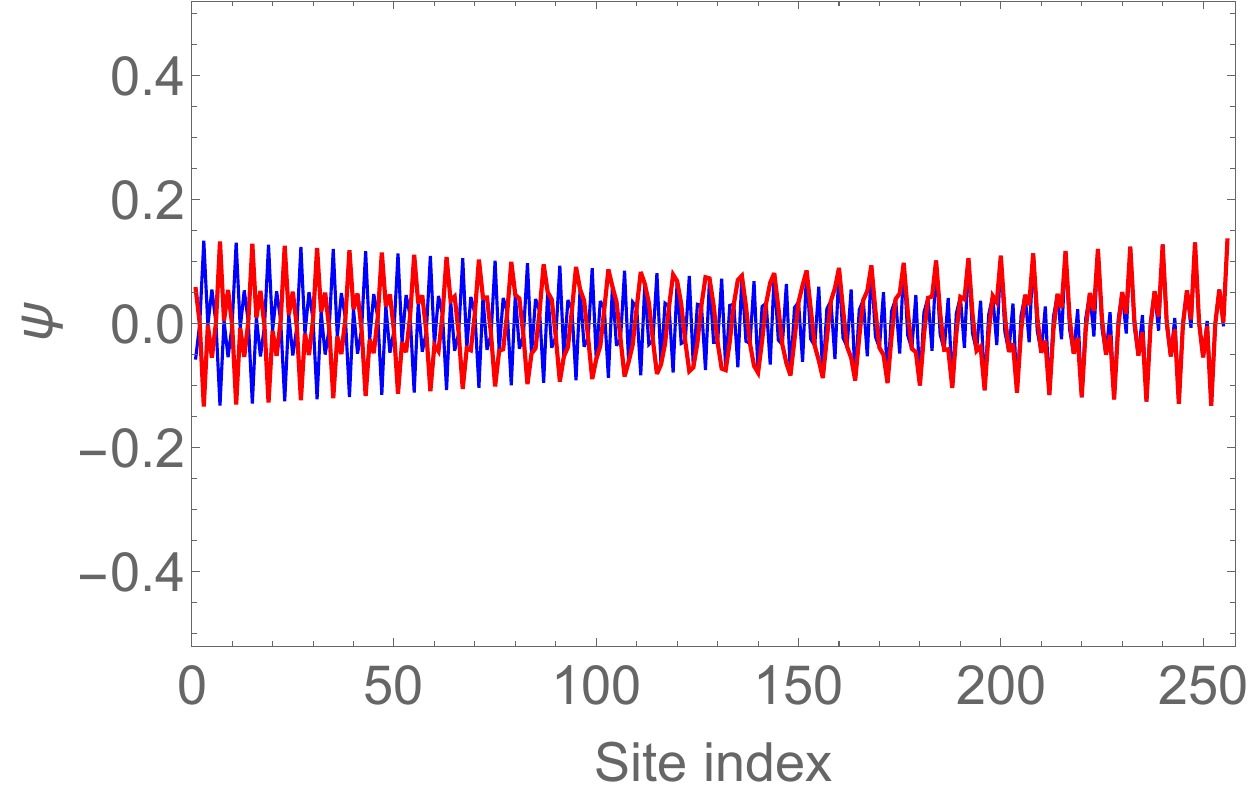}
  \includegraphics[width=.45\linewidth, height=1.25 in]{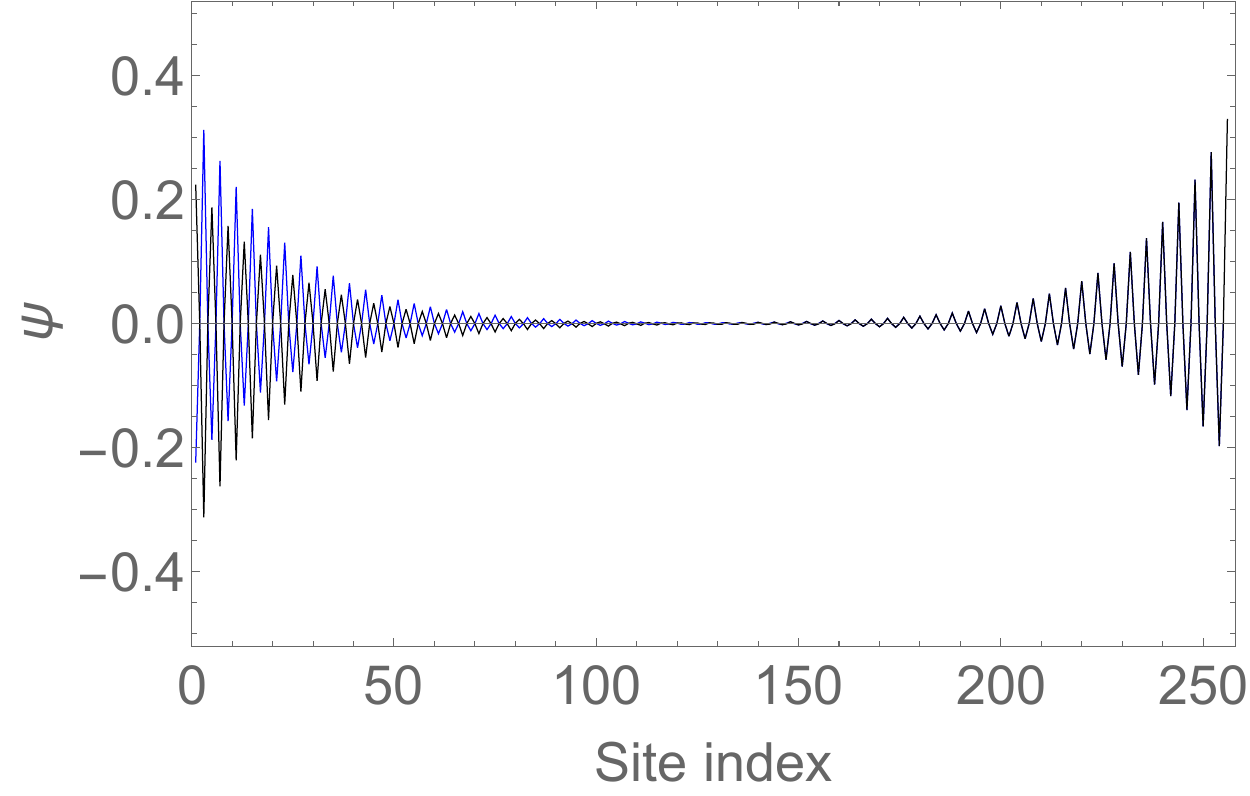}}
     \put(-10,62){(a)}
     \put(125,58){(b)}
    \end{picture}\\
     \vskip -.00004 in
   \begin{picture}(100,100)
     \put(-70,0){
    \includegraphics[width=.45\linewidth,height=1.25 in]{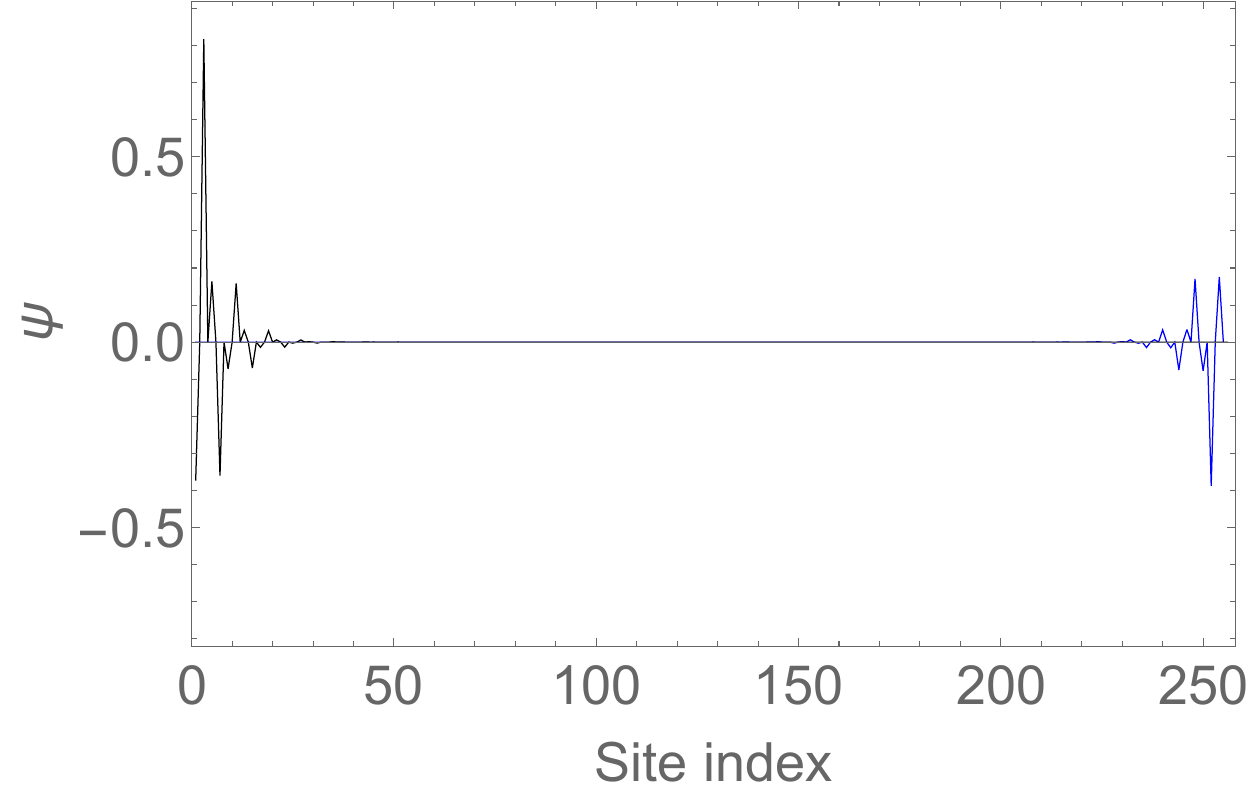}
    \includegraphics[width=.45\linewidth,height=1.25 in]{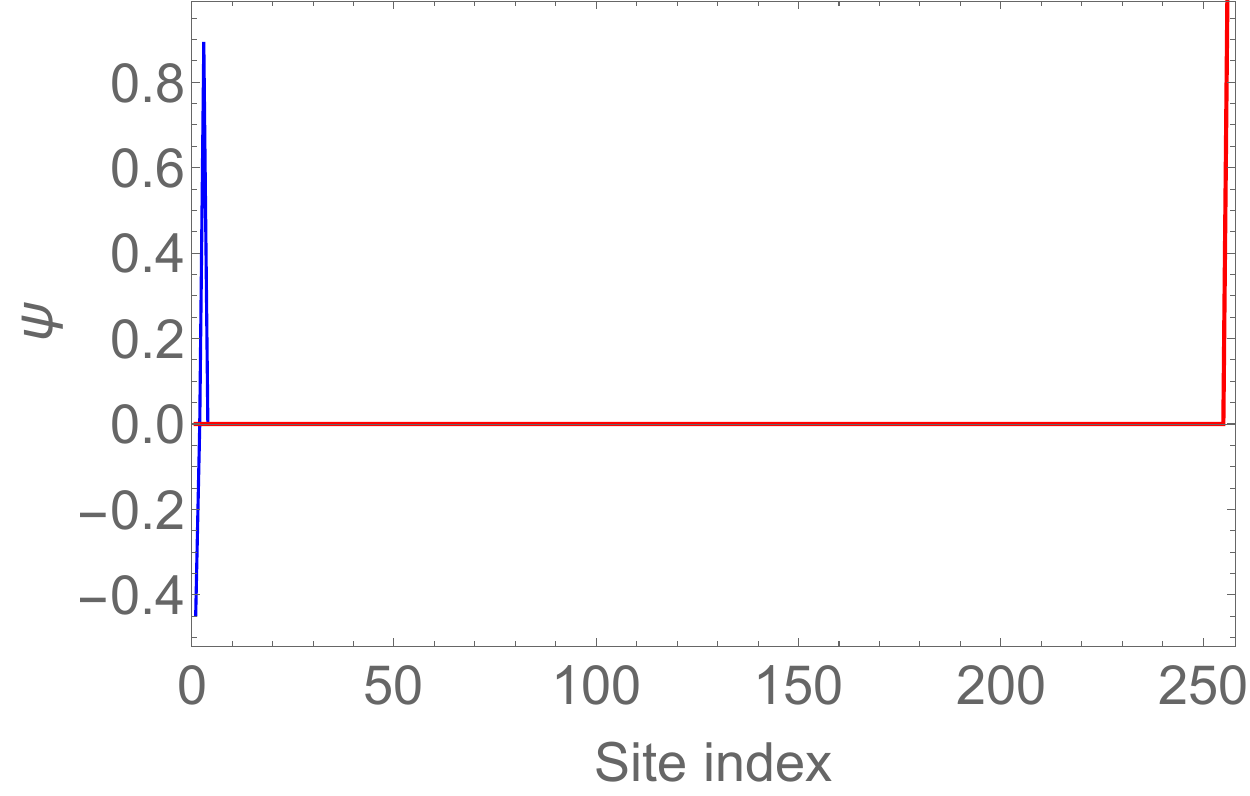}}
   \put(-10,62){(c)}
    \put(125,58){(d)}
   \end{picture}
  \vskip -0.1 in
\caption{ Behavior of edge modes for different values of $\Delta$ for $\theta=\pi/2$. In (a) $|\Delta|=0, \sqrt{2}t$, (b) $|\Delta/t|= 0.4$, (c) $|\Delta/t|=1.2$, and (d) $\Delta=1$.} 
\label{figedge}
\end{figure}

Now, we plot the behavior of the edge modes near the critical, metallic-like and fixed points for $\theta=\pi/2$ in Fig. \ref{figedge}. From Fig. \ref{figedge}(a), one can notice a complete delocalization of the edge modes at the critical or metallic-like points $|\Delta|=0,\sqrt{2}t$. However, we get an increasing localization of the edge modes as we move towards the stable fixed points $\Delta=\pm t$. Interestingly, the edge modes decay faster in the bulk near the $|\Delta/t|=\sqrt{2}$ point compared to near the $\Delta=0$ point (see Figs. \ref{figedge}(b) and (c)). However, they exhibit maximum localization at the stable fixed points as shown in Fig. \ref{figedge}(d).


\vskip 2 in

\begin{thebibliography}{00}
\bibitem{ludwig}  A. P. Schnyder, S. Ryu, A. Furusaki and A. W. W. Ludwig, ``Classification of topological insulators and superconductors in three spatial dimensions'', Phys. Rev. B {\bf 78}, 195125 (2008).
\bibitem{kitaev} A. Kitaev, V. Lebedev and M. Feigel’man, ``Periodic table for topological insulators and superconductors'', None 1134, 22 (2009).
  \bibitem{chen0} P. Molignini, R. Chitra and W. Chen, ``Unifying topological phase transitions in non-interacting, interacting, and periodically driven systems, Europhys. Lett. {\bf 128}, 36001 (2020).
  \bibitem{rg}W. Chen, ``Scaling theory of topological phase transitions'', \blue{J. Phys.: Condens. Matter {\bf 28}, 055601 (2016).}
 \bibitem{rao}F. Abdulla, P. Mohan, and S. Rao, "Curvature function renormalization, topological phase transitions, and multicriticality``, \blue{Phys. Rev. B {\bf 102}, 235129 (2020).}
\bibitem{rg5}W. Chen, Weakly interacting topological insulators: Quantum criticality and the renormalization group approach, \blue{Phys. Rev. B {\bf 97}, 115130 (2018).}
\bibitem{rg2}P. Molignini, W. Chen, and R. Chitra, ``Generating quantum multicriticality in topological insulators by periodic driving", \blue{Phys. Rev. B {\bf 101}, 165106 (2020).}
\bibitem{rg1}P. Molignini, W. Chen, and R. Chitra, ``Universal quantum criticality in static and Floquet-Majorana chains", \blue{Phys. Rev. B {\bf 98}, 125129 (2018).}
\bibitem{rg3} W. Chen and A. P. Schnyder, \blue{New J. Phys. {\bf 21}, 073003 (2019).}
\bibitem{rg4}W. Chen, M. Sigrist, and A. P. Schnyder, \blue{J. Phys.: Condens. Matter {\bf 28}, 365501 (2016).}
\bibitem{rg6}E. P. L. van Nieuwenburg, A. P. Schnyder, and W. Chen, \blue{Phys. Rev. B {\bf 97}, 155151 (2018).}
\bibitem{edge1}W. Chen, M. Legner, A. Rüegg, and Manfred Sigrist, Correlation length, universality classes, and scaling laws associated with topological phase transitions, \blue{Phys. Rev. B {\bf 95}, 075116 (2017)}.
\bibitem{rg7}S. Kourtis, T. Neupert, C.Mudry, M.Sigrist, and W.Chen, \blue{Phys. Rev. B {\bf 96}, 205117 (2017).}
\bibitem{ssh}W. P. Su, J. R. Schrieffer, and A. J. Heeger, ``Solitons in Polyacetylene", \blue{Phys. Rev. Lett. {\bf 42}, 1698 (1979).}
\bibitem{wall2}W. P. Su, J. R. Schrieffer, and A. J. Heeger, “Soliton excitations in polyacetylene”, \textcolor{blue}{Phys. Rev. B {\bf 22}, 2099 (1980).}
\bibitem{kar}S. Kar, Edge state behavior in a Su–Schrieffer–Heeger like model with periodically modulated hopping, \blue{J. Phys.: Condens. Matter {\bf 36}, 065301 (9pp) (2024)}.
\bibitem{mandal}S. Mandal, S. Kar, Topological solitons in a Su-Schrieffer-Heeger chain with periodic hopping modulation, domain wall, and disorder,	\blue{Phys. Rev. B {\bf 109}, 195124 (2024).}
\bibitem{s1}C.-K. Chiu, J. C. Y. Teo, A. P. Schnyder, and S. Ryu, Classification of topological quantum matter with symmetries, \blue{Rev. Mod. Phys. {\bf 88}, 035005 (2016).}
\bibitem{altand}A. Altland and M. R. Zirnbauer, Novel symmetry classes in mesoscopic normal-superconducting hybrid structures, \blue{Phys.Rev.B {\bf 55}, 1142 (1997).}
\bibitem{comment}The sign of $\Delta$ is insignificant for the energy spectrum while the sign changing, and the related band inversion is considerable for the wave functions, and the topological properties.
\bibitem{bernevig}B. A. Bernevig and T. L. X. Hughes, Topological Insulators and Topological Superconductors (Princeton University Press, Cambridge, 2013).
\bibitem{w1}P. Matveeva, T. Hewitt, D. Liu, K. Reddy, D. Gutman, and S. T. Carr, One-dimensional noninteracting topological insulators with chiral symmetry, \blue{Phys. Rev. B {\bf 107}, 075422 (2023).}
 \bibitem{w2}O. Balabanov, D. Erkensten, and H. Johannesson, Topology of critical chiral phases: Multiband insulators and superconductors, \blue{Phys. Rev. Res. {\bf 3}, 043048 (2021).}
 \bibitem{symmetry} Y. Niu, S. B. Chung, C.-H. Hsu, I. Mandal, S. Raghu, and S. Chakravarty, \blue{Phys. Rev. B {\bf 85}, 035110 (2012).}
\bibitem{symmetry1}W. DeGottardi, M. Thakurathi, S. Vishveshwara, and D. Sen, \blue{Phys. Rev. B {\bf 88}, 165111 (2013).}
\bibitem{s2}V. Gurarie, ``Single-particle Green’s functions and interacting topological insulators", \textcolor{blue}{Phys. Rev. B {\bf 83}, 085426 (2011).}
\bibitem{s3}Z. Wang and S. C. Zhang, ``Topological Invariants and Ground-State Wave functions of Topological Insulators on a Torus", \textcolor{blue}{Phys. Rev. X {\bf 4} 011006 (2014).}
\bibitem{s3a}Y. He and C.-C. Chien, ``Non-Hermitian generalizations of extended Su–Schrieffer–Heeger models", \textcolor{blue}{J. Phys.: Condens. Matter 33 085501 (2021).}
\bibitem{s1a}M. McIntyre and G. Cairns, “A new formula for winding number”, \textcolor{blue}{In: Geometriae Dedicata 46.2 (1993), pp. 149–159.}
\end{thebibliography}
\end{document}